\documentclass[%
 reprint,
superscriptaddress,
 amsmath,amssymb,
 aps,
longbibliography,
]{revtex4-2}

\usepackage{graphicx}
\usepackage{dcolumn}
\usepackage{bm}
\usepackage[caption=false]{subfig}
\usepackage{verbatim}
\usepackage{hyperref}
\usepackage{xcolor}
\usepackage[normalem]{ulem}
\usepackage{soul}
\hypersetup{colorlinks=true, linkcolor=blue, citecolor=blue, linktocpage}

\usepackage[capitalize]{cleveref}
\usepackage[caption=false]{subfig}
\newcommand{\phantomsubfloat}[1]{
    {
        \captionsetup[subfigure]{labelformat=empty}
        \subfloat[][]{#1}
    }
}

\begin{document}

\preprint{APS/123-QED}

\title{Dynamics of Symmetry-Protected Topological Matter on a Quantum Computer}

\author{Miguel Mercado}
\email{momercad@usc.edu}
\affiliation{Department of Physics and Astronomy, University of Southern California, Los Angeles, CA 90089-0484}

\author{Kyle Chen}
\affiliation{Pritzker School of Molecular Engineering, The University of Chicago, Chicago, IL 60637}

\author{Parth Hemant Darekar}%
\affiliation{Department of Electrical and Computer Engineering, University of Southern California, Los Angeles, CA 90089-0484}
\affiliation{Department of Electrical and Computer Engineering, University of Maryland, College Park, Maryland 20742}

\author{Aiichiro Nakano}
\affiliation{Collaboratory for Advanced Computing and Simulations, University of Southern California, Los Angeles, CA 90089-0484}

\author{Rosa Di Felice}
\affiliation{Department of Physics and Astronomy, University of Southern California, Los Angeles, CA 90089-0484}
\affiliation{CNR Institute of Nanoscience, Modena, Italy}

\author{Stephan Haas}
\affiliation{Department of Physics and Astronomy, University of Southern California, Los Angeles, CA 90089-0484}

\date{\today}

\begin{abstract}
Control of topological edge modes is desirable for encoding quantum information resiliently against external noise. Their implementation on quantum hardware, however, remains a long-standing problem due to current limitations of circuit depth and noise, which grows with the number of time steps. By utilizing recently developed constant-depth quantum circuits in which the circuit depth is independent of time, we demonstrate successful long-time dynamics simulation of bulk and surface modes in topological insulators on noisy intermediate-scale quantum (NISQ) processors, which exhibits robust signatures of localized topological modes. We further identify a class of one-dimensional topological Hamiltonians that can be readily simulated with NISQ hardware. Our results provide a pathway towards stable long-time implementation of topological quantum spin systems on present day quantum processors.

\end{abstract}

\maketitle

\emph{Introduction.}\textemdash Understanding topological properties of matter is a current frontier of physics \cite{RevModPhys.88.035005, RevModPhys.91.015006, topomechanics, PhysRevLett.114.114301, nano12193522, Sharma_2022, Kumar_2020}. Chiral phenomena, in particular, have been ubiquitously sought after in recent years for their novel emergent behaviors which have been observed, for example, in acoustic and mechanical systems \cite{Xue_2022, Zhang_2018, Serra_Garcia_2018, PMID:30598540, PhysRevLett.122.204301, PhysRevB.101.094107, PhysRevApplied.14.014084}, photonics \cite{Price_2022, El_Hassan_2019, PhysRevB.98.205147, PhysRevLett.122.233902, PhysRevResearch.2.013121}, and magnetic materials \cite{Bernevig_2022, Li_2019, PhysRevApplied.13.064058, PhysRevB.101.184404, 2021JAP...129o1101W, Zhuo_2023}. Symmetry-protected topological (SPT) phases, the paradigmatic models for realizing such chiral states, are characterized by their protection via global symmetries which have become well-studied in the space-domain \cite{Wen:1989iv, Ryu_2010, Qi_2011, Li_2021, Senthil_2015}. A consequence of topological symmetry protection is the predicted robustness of SPT phases against local noise and perturbations, opening avenues for foundational research and numerous device applications in quantum science \cite{Kitaev_2001, PhysRevB.85.075125, McClarty_2022, 2022NatMa..21...15H, osti_1992388}, such as novel high-performance transistors \cite{hu_2017, chen_2021, Vandenberghe_2017}, quantum sensors \cite{Budich_2020, Viti_2016}, and protected room-temperature transport devices \cite{Ohnishi_2022, Shumiya_2022, PhysRevLett.113.096601, Luo_2023}.

Despite this enticing context, thorough investigation of chiral topological modes within the time-domain curiously remains ambiguous, especially within the context of open system time evolution \cite{de_Groot_2022, PhysRevB.107.L241104}. Understanding the role of topology in the dynamics of quantum systems therefore, is a timely endeavor, which - however - has been exceptionally challenging to realize and control due to the inherent difficulty of accessing sensitive many-body quantum states \cite{PhysRevB.106.035414, PhysRevLett.114.017001, PhysRevX.10.041038}. 

As intrinsically quantum platforms, a pioneering application of quantum computers is the physical simulation of many-body systems \cite{Feynman1982-FEYSPW, 2012NatPh...8..264C, RevModPhys.86.153, Bassman_2021, Smith_2019}, where in special-use cases practical quantum advantage has recently been shown to be achievable \cite{Daley2022PracticalQA}. Such is the potential of quantum simulation that topological physics is currently being explored using photonic simulators \cite{Kitagawa_2012, PhysRevLett.122.173901, PhysRevLett.100.013904, Li:15, 2017NaPho..11..763K}, ultracold atoms \cite{de_L_s_leuc_2019, Ebadi_2021, cardarelli2022accessing, Zhang_2018, PMID:35228521, PhysRevLett.119.123601},  hybrid-quantum simulation \cite{Shi_2023}, and periodically driven quantum simulators \cite{Dumitrescu_2022, chen2020digital, Harle_2023}. 

While these provide a promising route to investigate condensed matter systems, many-body interactions that would otherwise be intractable for analog quantum simulators are well suited for digital quantum simulation, which provides universal capability to realize any finite-dimensional local Hamiltonian using sequences of quantum circuits \cite{Smith_2019, doi:10.1126/science.273.5278.1073, Satzinger_2021}. Although topological states have been prepared and observed using digital quantum simulation \cite{PhysRevLett.125.160503, Azses_2020, Choo_2018, koh_2022, Koh_2022_2, tan2021realizing, smith_2022}, the analysis of  transient behavior of topological modes in the time-domain presents a particular set of challenges. A primary reason is the effect of read-out noise and gate-error rates, which have sharply limited reliable quantum simulation of interacting many-body systems \cite{PhysRevB.101.184305}. 

The computational power and effectiveness of current noisy intermediate-scale quantum (NISQ) computers is restricted by 
the number and quality of qubits. In fact, NISQ-era qubits have short coherence times and high gate-error rates \cite{Preskill_2018}, which demand the requirement for noise mitigation and error correction schemes. Moreover, quantum simulations for generic Hamiltonians require that the circuit depth scales linearly at minimum with the number of simulation time steps, according to the No-Fast-Forwarding Theorem \cite{10.5555/2011373.2011380, Berry_2006, Atia_2017}. These restraints make prolonged time evolution of many-body systems exceedingly challenging for conventional approaches. 

To date, a reliable method to realize the long-time dynamics of topological phases of matter using digital quantum simulation is not available due to the aforementioned limitations. Prior simulations are largely restricted in the number and width of time-step intervals, making the analysis of transient dynamics, which is non-negligible when interaction with the environment is present, unreachable. 

We introduce a method to realize the long-time quantum dynamics of topological matter using contemporary quantum hardware, positioning NISQ computation as a favorable environment to probe novel phases of matter. In comparison to previous methods \cite{koh_2022, Koh_2022_2, PhysRevLett.125.160503, Azses_2020, Choo_2018, tan2021realizing, smith_2022}, which have been limited in the total time duration or restricted entirely to the space-domain, our protocol enables the quantum simulation of topological states in one spatial dimension nominally up to arbitrarily long times. We call to attention a subset of Hamiltonians with novel topological properties that are now tractable with this result. Its ability to predict physical results does not rely on employing readout error mitigation or post-processing techniques, though the output precision can be complemented by such techniques. By engineering interactions between qubits, we implement and analyze three different topologically non-trivial states. Specifically, we construct the desired topological phases via staggered coupling, using a topological mirror, and introducing a topological defect. We utilize the time-step compression of recently developed constant-depth quantum circuits \cite{Bassman_Oftelie_2022, PhysRevA.105.032420}, which possess the key property that the depth required to execute a quantum simulation is constant with respect to the number of time steps compared to linear or exponential. This inherent feature of matchgates enables the construction of topological Hamiltonians as quantum circuits to be simulated for arbitrary time. This is achieved because the constituents of constant-depth circuits, matchgates, have the symmetric property that the product of two matchgates is itself a matchgate, allowing for the decomposition into native gates with only two CNOT gates, while conventionally three CNOT gates are needed. This reduction enables the Trotter error to be made negligible by breaking the simulation down into small time steps. We point out that while fixed-depth circuits were recently designed to map spin and fermionic models onto quantum hardware \cite{PhysRevA.105.032420, PhysRevLett.129.070501}, their physical implementation towards  topological models and rigorous investigation of their experimental result has not yet been carried out. Subsequently, we provide an analysis of the long-time dynamics of bulk and edge modes, and observe that the topological modes are coherently stabilized in each case using our method, where transient data is clearly visible.  
    
\emph{Dynamics of topological insulators as constant-depth circuits.}\textemdash We simulate the time evolution of coupled SPT insulator systems using quantum circuits. In this approach, the non-trivial topology is introduced via locally varying nearest-neighbor couplings between lattice sites that host magnetic dipoles. 

\begin{figure}[hbt!]
\centering
    \includegraphics[scale = .50]{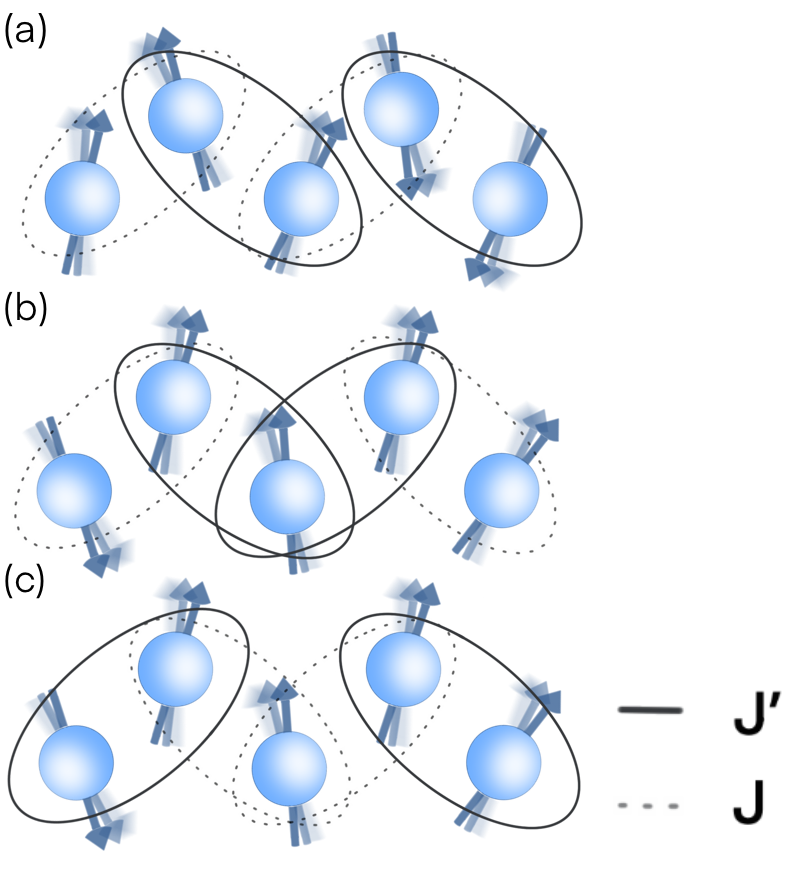}
    \vspace{-2\baselineskip}
    \phantomsubfloat{\label{fig:1a} }
    \phantomsubfloat{\label{fig:1b}}
    \phantomsubfloat{\label{fig:1c}}
    \vspace{-1\baselineskip}
    \caption{\label{fig:1} Schematic illustration of topological insulator models with alternating couplings $J$ and $J'$ ($> J$), realized in this work on quantum processors using five qubits. (a) Spin-1/2 chain with a topological surface mode observed on the first qubit. (b) Spin-1/2 chain with topological surface modes on the first and fifth qubits. (c) Spin-1/2 chain with a topological defect mode on the central qubit.}
\end{figure}

We implement the topological states using a 1D Quantum Spin Hamiltonian, given by

\begin{equation}
    \mathcal{H} = - \sum_{i=1}^{N-1} J_{z, i} Z_{i} Z_{i+1} -  h_{x}\sum_{i=1}^{N} X_{i}, 
     \label{eq:spin_hamiltonian}
\end{equation}

where $Z_{i}$, $X_{i}$ denote the spin-1/2 Pauli $z$ and $x$  matrices acting on qubit $i$, $J_{z,i}$ denotes the $z$-direction coupling between qubits $i$ and $i+1$, and $h_{x}$ is an external magnetic field in the $x$ direction. By particular choices of $J_{z,i}$, the signature bulk and edge modes of topological insulator systems emerge and are tunable; we extensively examine multiple cases. We note that while Eq. (\ref{eq:spin_hamiltonian}) presents a quantum Hamiltonian with couplings between qubits along the $z$ direction and external field in the $x$ direction, the model generalizes to couplings and fields along any direction, where nearest neighbor interactions are confined to two or less directions, and the external field acts along one perpendicular direction.

A schematic of the topological configurations is given in Fig. \ref{fig:1}. In general, when the coupling $J'$ is greater than $J$, the system enters a topologically nontrivial regime, and exhibits protected topological modes that exist on the weakly-coupled edges of the system \cite{Chen_2011, Pollmann_2010, PhysRevLett.42.1698, PhysRevD.13.3398, Meier_2016, Ryu_2010, PhysRevLett.109.096403, PhysRevB.104.125418, PhysRevB.102.235411}. Fig. \ref{fig:1a} shows a topological edge mode, which can be created and protected via alternating weak couplings $J$ and strong couplings $J'$ in a chain. Fig. \ref{fig:1b} shows  a  mirror configuration, which can be created by using weak couplings $J$ connecting the qubits on the edge, and strong couplings $J'$ connecting the ``bulk'' qubits, thus creating edge modes at the two ends of the chain. Fig. \ref{fig:1c} illustrates a topological defect, which can be created in the center of the chain by choosing strong couplings $J'$ at the edges and weak couplings $J$ connecting the bulk qubits. 

\begin{figure}
    \centering
    \includegraphics[scale=.40] {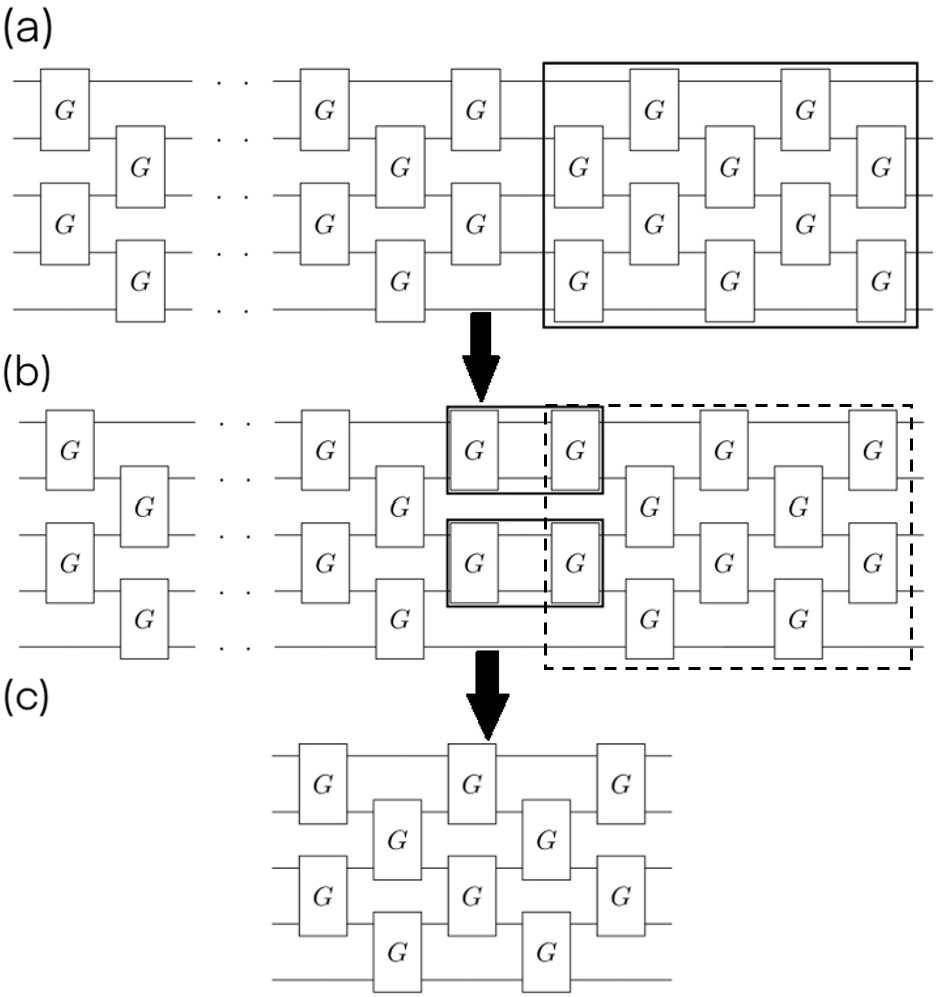}
    \vspace{-2\baselineskip}
    \phantomsubfloat{\label{fig:2a}}
    \phantomsubfloat{\label{fig:2b}}
    \phantomsubfloat{\label{fig:2c}}
    \caption{Building constant depth circuits for a 5-qubit system. (a) Matchgate symmetry is enforced on a set of matchgates outlined in bold. (b) Pairs of matchgates are mapped to singles using the SU(2) downfolding property of matchgates. Process is repeated for remaining matchgates in each set as denoted by the dotted line. (c) New circuit with number of matchgate columns reduced by one is produced.}\label{fig:2}
    
\end{figure}

\begin{figure*}[t]
    \centering
    \includegraphics[width=.85 \textwidth]{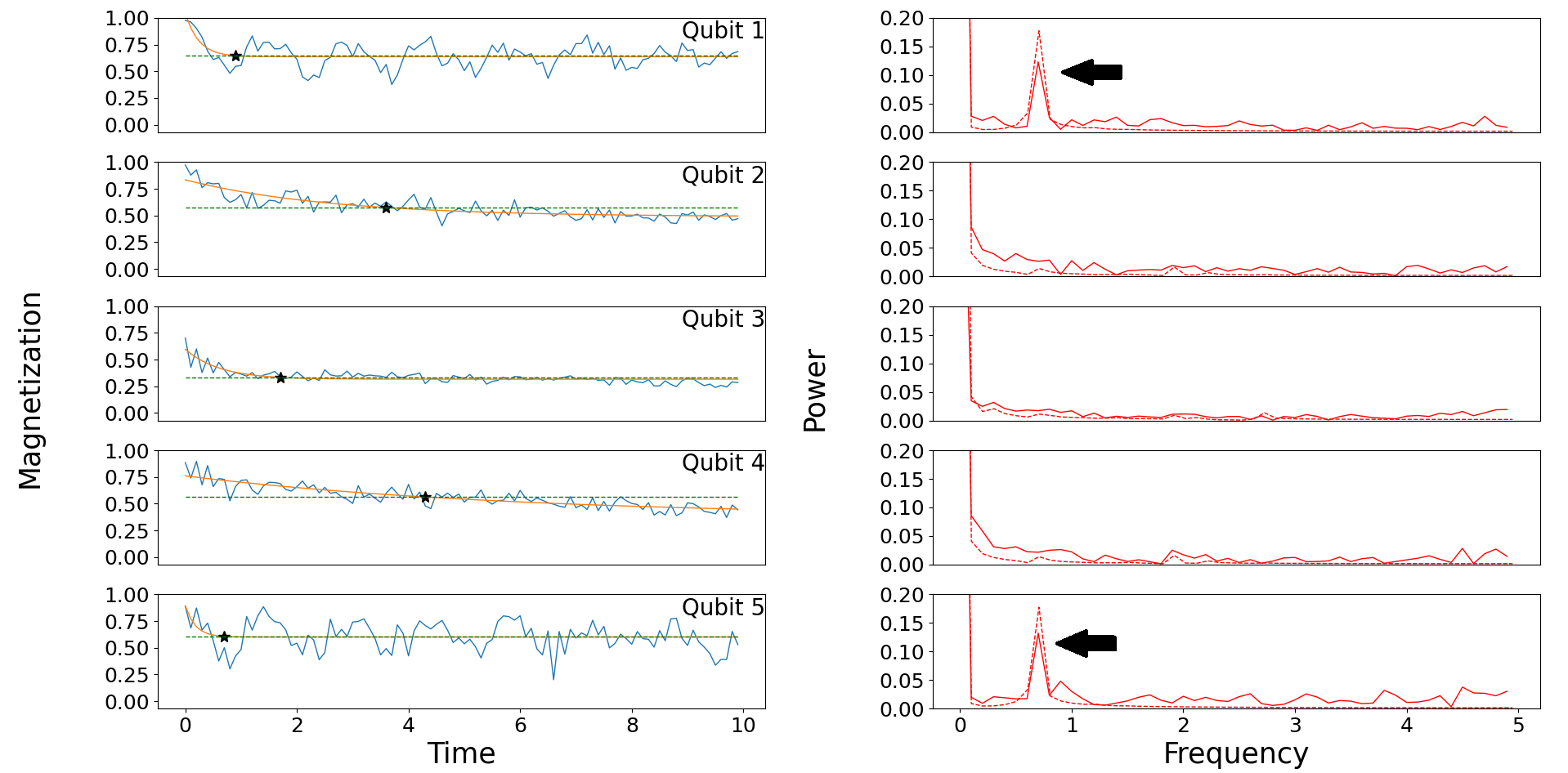}
    \caption{\label{fig:3}Local dynamics of a 5-qubit Hamiltonian in a topologically non-trivial regime, illustrated in Fig. \ref{fig:1b}, obtained by NISQ processors. $J = 2$, $J' = 4$, $h_x = 1$. Left: real-space magnetization in the $z$-direction for each qubit in solid blue lines. Transient dynamics is highlighted with orange lines which are fitted according to the lifetime damping formula exp($- \gamma t$). We report the values of the damping constant $\gamma$ as follows, qubit 1: 4.1324, qubit 2: 0.3843, qubit 3: 1.6278, qubit 4: 0.1686, qubit 5: 5.5488 in units of $h_{x}$. Black stars indicate the noisy finite system equilibration times for each qubit (defined in Sec. IIID of the Supplementary Material \cite{supp}), where the dotted green lines indicate the value of the local average magnetization of each qubit over the entire time evolution $t=0$ to $t=10$ (qubit 1: 0.6488, qubit 2: 0.5772, qubit 3: 0.3348, qubit 4: 0.5641, qubit 5: 0.6080 in units of $h_{x}$). Time is reported in units of the inverse magnetic field $1/h_{x}$. Right: Fourier-transformed power spectrum in frequency space (solid red lines). Sharp surface modes are clearly visible on the first and fifth qubits. Frequency is reported in units of the magnetic field $h_{x}$. Results are overlaid with its closed system power spectrum in dotted red lines (see Supplement Sec. IV \cite{supp} for complete closed system results for each model).}
\end{figure*} 

As a special class of quantum circuits, constant-depth circuits have a constant scaling rate in circuit depth with time step, as compared to conventional circuits or low-depth circuits which scale linearly or exponentially. This class of Hamiltonians thus embodies a nontrivial exception to the No-Fast-Forwarding Theorem \cite{Atia_2017, Gu_2021}. The time-step compression of constant-depth circuits follows directly from their intrinsic matchgate property, which utilizes symmetry to downfold the total amount of circuits needed to construct quantum algorithims on native gates.

According to this property, the product of two matchgates is also a matchgate, constituting an SU(2) Lie Algebra \cite{Bassman_Oftelie_2022}. This allows the decomposition of a matchgate into native gates, which conventionally requires three CNOT gates, to be executed with only two CNOT gates with constant-depth circuits. Therefore, $N(N-1)$ CNOT gates are required to simulate a system with $N$ spins. This compression in resource requirement allows the  Trotter error, which is proportional to the size of the time step, to be negligible, by breaking down the simulation into miniature time steps which can be repeated many times. We elaborate on this method in Sec. II of the Supplementary Material \cite{supp}.

In Fig. \ref{fig:2}, we illustrate the construction of constant-depth circuits for 5-qubit spin systems. We map SPT insulators into constant-depth circuits, splitting the total evolution into discretized time steps $\Delta t=0.1$ in units of the inverse applied magnetic field $\hbar/h_{x}$ following the convention $\hbar=1$. In principle, one can continually reduce the Trotter error by breaking down the total simulation time into increasingly smaller time-step intervals. 

In order to generate the circuits for these systems, we employ numerical optimization to identify optimal circuit parameters. This is accomplished by first computing the time evolution operator associated with each topological spin system, which defines a target matrix. After constructing the constant-depth circuit structure, which contains $N$ matchgate columns for a system with $N$ qubits, we compute its corresponding circuit matrix. Then, we generate optimized constant-depth circuits by minimizing the distance between the target and circuit matrix. This procedure is detailed in Sec. III of Supplementary Material \cite{supp}. We point out that generation of constant-depth circuits can be achieved independently of the optimization method used.

\emph{Realization on quantum hardware.}\textemdash We implement the time-evolution of SPT insulator models using an IBM transmon-based quantum computer \cite{gadi_aleksandrowicz_2019_2562111}, demonstrating that topological physics can be coherently observed with \emph{ibmq\_manila}, a quantum processor with five qubits. Our experiments using five-qubit chains illustrate that the strength of topological effects in the many-body dynamics are predominantly dictated by the contrast between J and J’. Because of their intrinsic gap, the equilibration time of each topological model is only weakly dependent on system size \cite{PhysRevA.87.032108}. Moreover, we expect exponentially fast convergence in system length of gapped states \cite{PhysRevB.45.9798}; these two observations have been checked empirically using additional hardware calculations. As realized on NISQ processors, each model is subjected to environmental noise that is intrinsic to the quantum device itself \cite{PRXQuantum.4.040329}, which exists independently from the noise of Trotter error-type which is mitigated by constant-depth circuits. We compare our results with classical simulations, using Exact Diagonalization to numerically solve the closed-system dynamics as ground truth (Sec. IV of the Supplementary Material \cite{supp}).

The IBM hardware results for the topological mirror case in Fig. \ref{fig:1b} are shown in Fig. \ref{fig:3}, where the $z$-direction magnetization is reported for each qubit as a function of time, over the total open system evolution. Presented in the left panel of Fig. \ref{fig:3} is the real-space data obtained from the quantum hardware solid blue lines), while the right panel of Fig. \ref{fig:3} shows the corresponding power spectrum in frequency space generated by performing a Fourier Transform solid red lines. The experimental result is plotted against its respective closed system Fourier spectrum in dotted red lines for reference. 

For this configuration, we expect strongly localized topological surface modes on both open ends of the quantum spin chain. Furthermore, since there is spatial mirror symmetry in the system parameters, qubits 1 and 5 should be equivalent, as well as qubits 2 and 4. 

Indeed, the result obtained from the quantum hardware accurately corroborates these expectations. A clear signature of topological surface modes is evidenced by the strong temporal oscillations of the magnetization in qubits 1 and 5, translating into sharp low energy modes in the power spectrum, indicated by arrows in Fig. \ref{fig:3}. Their frequency depends on the choice of Hamiltonian parameters $J, J', h_x$, whereas their lifetime is characterized by the damping rate $\gamma$ as in a fitting formula, $m(t) \propto \exp(-\gamma t)$. We remark that in the complementary closed system calculation, this frequency is equivalent to the energy difference between pairs of eigenstates. These topological mode signatures are the salient feature preserved amidst noise in the hardware results. Constant-depth circuits have stabilized the SPT modes, producing longer-lived states at the edge of the system, which has not been observable with conventional quantum circuits (see Supplementary Fig. S13 \cite{supp}). 

In contrast, the``bulk" qubits 2, 3, and 4 do not display any oscillations, in analogy with the equivalent closed quantum system dynamics \cite{PhysRevB.106.035414} discussed in Sec. IV of the Supplementary Material \cite{supp}. Furthermore, we observe site dependent transient dynamics in the time window $t\in [0,4]$, with the odd site qubits settling into their steady state markedly earlier than the even sites, as indicated by the black dots in the left panel of Fig. 3, which were obtained using the criteria for equilibration times of noisy finite systems outlined in Ref. \cite{PhysRevA.87.032108}, namely, when the expectation value of an observable approaches its average $\overline{\mathcal{A}}$ and begins fluctuating about it with fluctuations $\delta\mathcal{A} = \sqrt{\overline{\mathcal{A}^{2}} - \overline{\mathcal{A}}^{2}}$. Here we choose a definition based on the physical reasoning that finite sized, gapped quantum states produce quasi-periodic dynamics, which is visible in the local magnetization. The orange lines are obtained by fitting the exponential damping function to the short-time data, indicating with the black star where this transient functional dependence does not apply anymore. The corresponding value of the local average magnetization $\overline{\langle \sigma^{z}_{i} \rangle} = \frac{1}{T}\int_{0}^{T}\langle \sigma^{z}_{i}(t) \rangle dt$ is plotted as the dotted green line. We conclude that the topological surface modes equilibrate rapidly and fluctuate robustly upon reaching their steady state, in contrast to the bulk qubits which undergo gradual temporal decay and require going out to much longer times to observe their full equilibration---underscoring the stabilizing effect of topology in the open quantum system dynamics. Finally, the observed large DC (zero energy) feature in the power spectrum is due to the polarization of the qubit spins along the $z$-direction. 

Moving to the other topological configurations shown in Figs. \ref{fig:1a} and \ref{fig:1c}, we also find agreement between real-time evolution on NISQ hardware and the equivalent closed system quantum dynamics. Specifically, in the staggered coupling model of Fig. \ref{fig:1a}, a sharp amplitude in Fourier space is detected on the first qubit, which is weakly coupled with the bulk, indicating a long lived topological surface mode. In the topological defect case of Fig. \ref{fig:1c}, a strongly localized state with temporal low-frequency oscillation is observed on the center qubit, (Supplementary Material Sec. IIIB \cite{supp}). We add that the corresponding shot noise reported is a minor effect compared to the overall trend in the local magnetization, which is clear by isolating the small fluctuations from the local moving average of the magnetization of all qubits (see Supplementary Material Sec. IIID \cite{supp}).

We emphasize that the presented results are measured data on NISQ devices, without any noise mitigation or error correction techniques. Despite this, the constant-depth method measurably outperforms conventional Trotterization, in which each qubit undergoes rapid decay where noise due to large Trotter error scrambles discernable signatures in the data (see Supplementary Fig. S13 \cite{supp}). Using this method to isolate the Trotter error, our results show that topological phases exhibit robust dynamics when subjected to the intrinsic noise of quantum devices, which is unavoidable due to pervasive hardware defects or ambient interactions. Our study demonstrates that circuit compression techniques can be effectively used with available NISQ devices to study dynamics of novel phases of matter despite the known NISQ challenges of noise and limited system size. 

\emph{Fast-forwardable topological models.}\textemdash The results reported here are general and can be utilized to realize the long-time dynamics for a host of topological systems. We illustrate how to apply our method to probe topological phenomena in open systems where the environmental interaction is driven by the intrinsic noise of the gate-based quantum device and highlight a notable subset of models, pointing out experimentally accessible procedures to control non-trivial states. First, one can straightforwardly realize protected modes by manufacturing corresponding symmetries of a desired Hamiltonian $H$ through artificial design of couplings $J$, demonstrated by models depicted in Fig \ref{fig:1}. Particularly, the presence or absence of chiral symmetry $\mathcal{C}$ determines the existence of topological modes in SPT phases \cite{Ryu_2010, PhysRevB.102.235411, Liu_2023, Kitaev_2009, PhysRevB.55.1142}, which can be further categorized into chiral-symmetric Hamiltonians that maintain or break combinations of time-reversal $\mathcal{T}$ symmetry and particle-hole (charge-conjugation) $\mathcal{P}$ symmetry. It has been proven that quadratic Hamiltonians can be fast-forwarded \cite{Atia_2017}, meaning that a system with $N$ spins requires only $N(N-1)$ CNOT gates to simulate. This follows from the fact that matchgates for quadratic Hamiltonians can be decomposed into native circuits requiring two CNOT gates. For further technical details, see Supplementary Material Sec. II \cite{supp}. 

Therefore, one can completely tune the presence and character of topological modes through our method by modulating interactions $J$ on two or less axes to realize discrete symmetries in $\mathcal{H}$ of particular interest, such as chiral symmetry $\mathcal{C}$, which is then decomposed into matchgates. This can be written generally as $\mathcal{H} = c^{\dag} H c$, where
\begin{equation}
    H =  \begin{pmatrix}
    A & B &\\
    -B^{*} & -A^{*} &
     \end{pmatrix},
\end{equation}
with arbitrary matrices $A$, $B$ satisfying $A = A^{\dag}$ and $B^{T} = -B$, and $c^{\dagger}_{j}$ and $c_{j}$ denoting the fermionic creation and annihilation operators which define the column vector operator $c = (c_{1} \hdots c_{n} c^{\dag}_{1} \hdots c^{\dag}_{n})^{T}$. obeying
$U_{\mathcal{C}}^{-1} \mathcal{H} U_{\mathcal{C}} = -\mathcal{H}$,
where $U_{\mathcal{C}}$ is defined as the chiral symmetry operator,  $\mathcal{C} = \mathcal{P} \cdot \mathcal{T}$.

We argue that this opens many interesting directions in one dimension, particularly in studying open system dynamics of models in the AIII, BDI, CI, and DIII universality classes \cite{Kitaev_2009}. In higher dimensions, it has been shown that $n$-band topological insulator Hamiltonians can be mapped to 1D hardcore boson chains by representing their single-particle hopping as an $n$-particle hopping living on distinct sublattices of a 1D chain as shown in Ref. \cite{Koh_2022_2}, which can be applied to generalize our method through application of matchgate symmetries. We point out that it has recently been shown that certain number-conserving bosonic Hamiltonians can also be fast-forwarded \cite{Gu_2021}. Another procedure to probe topological modes is to induce spatial localization within the lattice by modulating $J$. This is exemplified by the model in Fig. \ref{fig:1c}, which realizes a topological defect which can be interpreted as a domain wall in the center of the system between two configurations \cite{Estarellas_2017, Teo_2017}, we further elaborate on this process in Sec. IC of the Supplementary Material \cite{supp}. 

\emph{Discussion.}\textemdash We introduced a method that greatly improves dynamical simulations of quantum systems using contemporary quantum processors. By utilizing the time-step compression of constant-depth circuits, the transient dynamics of novel many-body systems is achievable up to arbitrary long times, which to date has remained one of the major challenges of the NISQ era.

Using this approach, we showed how open system quantum dynamics where the environmental interaction is driven by the intrinsic noise of the quantum hardware itself can be stabilized by exploiting topological properties of matter. While conventional qubits suffer from rapid decoherence, one can create much longer lived qubits through utilizing the topologically stabilized surface states as shown in our study. These long lived states present a host of potential applications. For example, one may use such topologically stabilized states to create robust and highly local quantum sensing devices through tuning the magnetic field $h_{x}$ and couplings $J$ and $J'$ closer relative to each other to induce a critical state \cite{PhysRevLett.129.090503}. In this state, the qubits have higher sensitivity, which combined with the advantage of being longer-lived can potentially be applied as a quantum sensor. Alternatively, one can tune $h_x$ farther than $J$ and $J'$ to maximize the contrast between $J$ and $J'$. In principle, tuning the system as such will increase the effectiveness of using topological modes as qubits with longer-lived coherence times.

The constant-depth quantum circuits method provides a clear pathway towards executing long-time dynamical simulations of topological quantum spin systems. Our one-dimensional protocol can be extended to higher dimensions by complementing it with larger system implementations and mathematical developments to the regime of validity of the matchgate algebra \cite{kökcü2023algebraic}. While in this study we demonstrate the versatility of this approach by executing dynamics of simple paradigmatic topological models with relatively small resource requirements, one can also apply the method to study long-time dynamics of more complex cases, such as in Aubry-André-Harper models \cite{Aubry_1980, Harper_1955, PhysRevB.14.2239, PhysRevLett.61.2144, Biddle_2009, PhysRevB.41.5544}, generalized Su-Schrieffer-Heeger models \cite{Xie_2019, PhysRevB.101.195117, PhysRevLett.107.166804}, and Kitaev chains \cite{Kitaev_2001, PhysRevLett.100.096407}. We also identify the circuit implementation of integrable interacting and open systems as a possible future direction. We point out that generally while simulating systems with dissipation necessitates at minimum linear scaling, if the system is integrable, then compression techniques can be used to generate circuits with constant depth scaling for this special class of interacting systems. Another promising future direction is dynamical studies of Higher Order Topological Insulators (HOTI)s \cite{Schindler_2018, PhysRevB.107.045118, PhysRevLett.128.127601}, which can be made tractable with ongoing improvements in circuit compression methods.

\textbf{Acknowledgements}
We thank Lindsay Bassman Oftelie of CNR-NANO, Pisa, Italy for valuable discussion and insight. AN was supported by an NSF grant OAC-2118061. RDF was partially supported by the Defense Advanced Research Projects Agency (DARPA), Contract No. HR001122C0063, by PNRR-Italy HPC Spoke 10 and by PRIN 2022W9W423. We acknowledge the use of IBM Quantum services for this work, including devices available through Oak Ridge National Laboratory (QCUP CPH150). The views expressed are those of the authors, and do not reflect the official policy or position of IBM or the IBM Quantum team.

\nocite{*}
\nocite{PFEUTY197079, Senthil_2015, valiant_2002, Bogolyubov:1958kj, JACKIW1981253, Valatin:1958ja, PhysRevB.84.195452, Qi_2011}

\bibliography{main}

\end{document}


\preprint{APS/123-QED}

\title{Supplementary Material: Dynamics of Symmetry-Protected Topological Matter on a Quantum Computer}

\author{Miguel Mercado}
\email{momercad@usc.edu}
\affiliation{Department of Physics and Astronomy, University of Southern California, Los Angeles, CA 90089-0484}

\author{Kyle Chen}
\affiliation{Pritzker School of Molecular Engineering, The University of Chicago, Chicago, IL 60637}

\author{Parth Hemant Darekar}%
\affiliation{Department of Electrical and Computer Engineering, University of Southern California, Los Angeles, CA 90089-0484}
\affiliation{Department of Electrical and Computer Engineering, University of Maryland, College Park, Maryland 20742}

\author{Aiichiro Nakano}
\affiliation{Collaboratory for Advanced Computing and Simulations, University of Southern California, Los Angeles, CA 90089-0484}

\author{Rosa Di Felice}
\affiliation{Department of Physics and Astronomy, University of Southern California, Los Angeles, CA 90089-0484}
\affiliation{CNR Institute of Nanoscience, Modena, Italy}

\author{Stephan Haas}
\affiliation{Department of Physics and Astronomy, University of Southern California, Los Angeles, CA 90089-0484}

\date{\today}

\begin{abstract}

\end{abstract}

\maketitle


\section{Topological Lattice Models}\label{sec:topo_models}
Here, we discuss the salient  features of the models depicted in Figs. 1(a)-(c) of the main text and consider straightforward methods to realize non-trivial topological degeneracy in interacting many-body systems. 

\subsection{Staggered Coupling}\label{sec:staggered}
First, we design an N-site spin-1/2 molecule in analogy to the  fermionic Su-Schrieffer Heeger model (SSH) \cite{PhysRevLett.42.1698}, as depicted in Fig. 1(a) of the main text. Since their inception, symmetry-protected topological (SPT) states of matter \cite{Wen:1989iv, Qi_2011, Ryu_2010, Senthil_2015, Gooth_2023, Li_2021} have become emblematic due to their topological surface modes \cite{JACKIW1981253} which emerge in the presence of one or more global symmetries of the Hamiltonian. Physically, the soliton states of SPT phases can be interpreted as boundary states between regimes with different ground states whose eigenenergies are within a gap between two bulk energy bands \cite{Estarellas_2017, PhysRevB.84.195452}. In the 1d SSH model, the global chiral symmetry $\mathcal{C}$ of the Hamiltonian plays this role, resulting in protected surface states that reside at zero energy  within an energy gap \cite{Ryu_2010}. 

The 1d SSH model provides an intuitive picture for realizing such chiral states via dimerization, i.e., staggered couplings between two different sublattices of the system. We emulate the physics of the SSH model in an N-site spin-1/2 molecule by introducing intra-cell (weak) $J$ and inter-cell (strong) $J'$ couplings between unit cells, which is implemented via alternating interactions between the spin-1/2 sites,
\begin{equation}
    \mathcal{H} = - \sum_{i=1}^{N-1} J^{\mu}_{i} \sigma^{\mu}_{i} \sigma^{\mu}_{i+1} -  h^{\mu}\sum_{i=1}^{N} \sigma^{\mu}_{i}, 
    \label{eq:spin-chain}
\end{equation}
where $\mu$ represents axes which are specified along $\{x,y,z\}$, $\sigma^{\mu}_{i}$ are the $\mu$ Pauli matrices acting on site $i$, $J^{\mu}_{i}$ is the interaction between neighboring Pauli matrices along the axis $\mu$, and $h^{\mu}$ is an external field acting along the $\mu$ direction. One recovers the familiar Transverse-Field Ising (TFI) model when $J^{\mu}_{i}$ is chosen to be uniform along the entire lattice with an applied magnetic field $h^{\mu}$ acting in its transverse direction. We remark that Eq. \ref{eq:spin-chain} is not the description for a fully general spin chain that can be can implemented using the method presented in the main text. One may, for example, select interactions $J^{\mu}_{i}$ acting on two different axes. We discuss this in further detail in Sec. \ref{sec:cd_circuits_const} and Sec. \ref{sec:qhardware}. In our study, we select interactions along the $z$-axis $\{J^{z}_{i}, \sigma^{z}_{i}\}$ with an external magnetic field along the $x$-axis $\{h^{x}, \sigma^{x}_{i}\}$.

To create the model featured in Fig. 1(a) of the main text, we induce topological soliton states analogous to those of the SSH model by enforcing two distinct sublattices via dimerization in an open chain: $J^{z}_{1} = 2$, $J^{z}_{1} = 4$, $J^{z}_{3} = 2$, $J^{z}_{4} = 4$, $J^{z}_{5} = 0$, with $h^{x} = 1$, noting that the last term $J^{5}_{1} = 0$ is due to the open boundary condition (OBC) necessary to host chiral edge states in 1d. To make our notation more intuitive, we define 

\[ \begin{cases} 
      J = J^{z}_{i} &\text{when } J^{z}_{i} = 2 \\
      J' = J^{z}_{i} &\text{when } J^{z}_{i} = 4\\
   \end{cases}
\]

to recover the familiar weak couplings $J$ and strong couplings $J'$ depicted in Fig. 1 of the main text. Spatially localized surface states are sharpened by increasing the contrast between $J$ and $J'$. 

This spin model has two distinct sublattices which can be viewed as an analogous version of the fermionic intra-cell and inter-cell hopping amplitudes between each unit cell. When the inter-cell hopping amplitude between two unit cells overcomes the intra-cell hopping amplitudes between two spins in one unit cell, our model gives rise to a bound soliton spin state at zero energy along the weakly-coupled edge of the spin lattice which is supported by the closed dynamics result discussed in Sec. IV. 
This state is localized within the gap between bulk energy eigenstates and is therefore protected against local perturbations such as the observed environmental noise affecting the bulk qubits; in principle the topological degeneracy of the system would be affected if environmental noise affects the hopping amplitudes in such a manner that the intra-cell hopping $J$ overcomes the inter-cell $J'$, which would close the energy gap and hybridize the surface and bulk states---which according to Fig. 3 of the main text, is not observed. Interestingly, our result demonstrates that the effect of spatial localization induced by sublattice dimerization is visible within the time-evolution of a system with minimal lattice length, which was previously not well-studied.

We now briefly discuss the transverse magnetic field dependence. The ferromagnetic phase of the spin chain $J > h_x$ preserves the long-order (two-fold degeneracy) of the ground state, and thus the topological soliton zero state of the staggered coupling is coherently observed in this regime, hence why we choose $h_{x} = 1$. We note that increasing the magnetic field $J << h_x$ lifts this degeneracy, and thus soliton states that originally result as a consequence of the topological degeneracy of the model hence become indistinguishable during simulation,  consistent with our control hardware tests. 

Finally,  it is useful to analyze the topological modes from the viewpoint of symmetry as discussed in the main text. Topological protection of SPT phases arises from enforcing global symmetries of the Hamiltonian. The SSH-inspired molecule in this case has chiral (sublattice) symmetry $U_{\mathcal{C}}^{-1} \mathcal{H} U_{\mathcal{C}} = -\mathcal{H}$, which is responsible for the large degree of observed spatial localization of the topological mode on qubit 1 of the staggered coupling model limit of Eq. \ref{eq:spin-chain}. However, one may for example, design other global symmetries such as combinations of time-reversal and charge-conjugation symmetries, or crystalline symmetries \cite{RevModPhys.88.035005}. Spin-1/2 models such as the one presented in the main text can be related to fermionic/hardcore bosonic models through the non-local Jordan-Wigner transformation,

\begin{align}
    \sigma^{x}_{i} &= \prod^{i-1}_{j}(1-2n_{j})(c^{\dag}_{i} + c_{i}) &
    \sigma^{y}_{i} &= \prod^{i-1}_{j}(1-2n_{j})(c^{\dag}_{i} - c_{i}) &
    \sigma^{z}_{i} &= (1-2n_{i}) &
    \label{eq:JW}
\end{align}

where $c^{\dag}_{i}$, $c_{i}$ denote the usual fermionic creation and annhilation operators and $n_{i} = c^{\dag}_{i}c_{i}$ the fermionic number operator acting on site i. We point out that the SSH model and Kitaev chain \cite{Kitaev_2001} obtained from applying the Jordan-Wigner transformation to the TFI model fall within the same universality class \cite{Lin_2017}, which is another way of viewing how the presented model exhibits robust edge modes on weakly-coupled boundaries in the topologically non-trivial phase.

\subsection{Mirror Model}\label{sec:mirror}
Building from the model described in Sec. \ref{sec:staggered}, we investigate topologies with more complex eigenvalue degeneracy and symmetry-breaking patterns. The topological mirror model featured in the main text is depicted by Fig. 1(b). We define the mirror model $\mathcal{H}_{M}$ as a topologically non-trivial limit of Eq. \ref{eq:spin-chain} with specified couplings $J^{z}_{1} = 2$, $J^{z}_{2} = 4$, $J^{z}_{3} = 4$, $J^{z}_{4} = 2$, $J^{z}_{5} = 0$, with $h^{x} = 1$. We discuss the spectrum of this model below. 

Consider $\mathcal{H}_{M}$ prepared in its initial state, before the quantum quench is performed, where $h_{x} = 0$. Let us examine the degeneracy of the initial phase of $\mathcal{H}_{M}$ when $(t < 0)$ to characterize the effect of enforcing the above couplings. Since $\mathcal{H}_{M}$ is a quadratic Hamiltonian, it is convenient to employ the fermionic Bogoliubov-de Gennes (BdG) approach to diagonalize $\mathcal{H}_{M}$ \cite{Bogolyubov:1958kj, Valatin:1958ja, PFEUTY197079}. 

Applying the Jordan-Wigner transformation \ref{eq:JW} to 
$\mathcal{H}$ 
gives the fermionic Hamiltonian with OBC,
\begin{equation}
\mathcal{H}^f =  - \sum^{N-1}_{i=1}( J^{z}_{i} c^{\dag}_{i} c_{i+1} + c^{\dag}_{i}c^{\dag}_{i+1} + h.c.) + h_{x}\sum^{N}_{i=1} (2c^{\dag}_{i}c_{i} - 1).
\end{equation}
Applying the same transformation \ref{eq:JW} to $\mathcal{H}_{M}$ 
gives the analogous fermionic Hamiltonian with OBC in the absence of magnetic field ($h_x=0$),
\begin{equation}
\mathcal{H}^f =  - \sum^{N-1}_{i=1}( J^{z}_{i} c^{\dag}_{i} c_{i+1} + c^{\dag}_{i}c^{\dag}_{i+1} + h.c.).
\end{equation}

Let $\Psi = (c_{1} \hdots c_{n} c^{\dag}_{1} \hdots c^{\dag}_{n})^{T} $,  $\Psi^{\dag} = (c^{\dag}_{1} \hdots c^{\dag}_{n} c_{1} \hdots c_{n})$ be the Nambu spinors which obey the usual fermionic anti-commutation relations. We can rewrite our problem into the standard BdG form as
\begin{equation}
    \langle\mathcal{H}_{M}^f \rangle= \Psi^{\dag} H \Psi = \Psi^{\dag} \begin{pmatrix}
    A & B &\\
    -B^{*} & -A^{*} &
     \end{pmatrix} \Psi \label{eq:BdG_M}
\end{equation}
where $H$ is the $2N \times 2N$ BdG block matrix containing matrices $A$ and $B$ satisfying $A = A^{\dag}$ and $B^{T} = -B$. It can be shown that in the case of the Ising model, $A$ and $B$ must be real and symmetric matrices whose elements yield the following structure for OBC \cite{PFEUTY197079}: 
\begin{align*}
A_{i,i} &= h^{x}& 
B_{i, i} &= 0&
A_{i, i+1} = A_{i+1, i} &= -\frac{J^{z}_{i}}{2}&
B_{i, i+1} = B_{i+1, i} &= -\frac{J^{z}_{i}}{2}& 
\end{align*}

Recalling that we examine the initial state where $h_{x}=0$, we obtain the following matrix
\begin{equation}
H = \frac{1}{2}\begin{pmatrix}
  \begin{matrix}
   & -J & & &  \\
   -J & 0 & -J' & &  \\
   & -J' & 0 & -J'&  \\
   &  & -J' & 0& -J  \\
   &  &  & -J & 0 \\
  \end{matrix}
  & \rvline & \begin{matrix}
    & -J &  & &  \\
   J & 0 & -J' & &  \\
   & J' & 0 & -J' &  \\
   &  & J' & 0& -J  \\
   &  &  & J & 0 \\
  \end{matrix} \\
\hline
  \begin{matrix} 
  & J &  & &  \\
   -J & 0 & J' & &  \\
   & -J' & 0 & J'&  \\
   &  & -J' & 0& J  \\
   &  &  & -J & 0 \\
  \end{matrix} & \rvline &
  \begin{matrix}
  & J &  & &  \\
   J & 0 & J' & &  \\
   & J' & 0 & J' &  \\
   &  & J' & 0& J  \\
   &  &  & J & 0 \\
  \end{matrix}
\end{pmatrix}
\end{equation}

The eigenvectors corresponding to the solution of Eq. \ref{eq:BdG_M} can be organized into a unitary matrix form 
\[\mathcal{U}_{M} =  \begin{pmatrix}
    U & V^{*} &\\
    V & U^{*} &
     \end{pmatrix}\label{eq:M_eigenvectors} \]
where the eigenvectors constitute the columns of the block matrices. To analyze the eigenspectrum obtained by solving Eq. \ref{eq:BdG_M}, define new Nambu fermion operators $\Phi = \mathcal{U_{M}}^{\dag}\Psi$, such that

\[\Phi  = \begin{pmatrix}
    U^{\dag} & V^{\dag} &\\
    V^{T} & U^{T} &
     \end{pmatrix}\begin{pmatrix} c_{1} \
    \hdots &
    c_{n}&
    c^{\dag}_{1}&
    \hdots &
    c^{\dag}_{n} \end{pmatrix}^{T} \]

The key feature of the $\mathcal{U}_{M}$ representation is to extract the Nambu fermion operator pairs that emerge from writing the linear combinations of the eigenvectors associated with each energy eigenvalue. Written in terms of Nambu fermion operators $\Phi$, we obtain the following spectrum of our mirror model:

\begin{equation*}
\begin{split}
\centering
& E_{1} = 8: \hspace{5mm}\Phi_{1} = -\frac{1}{\sqrt{2}}c_{3} -\frac{1}{2\sqrt{2}}(c^{\dag}_{2} - c_{2}) + \frac{1}{2\sqrt{2}}(c^{\dag}_{4} + c_{4}) \\
& E_{2} = 8: \hspace{5mm}\Phi_{2} = \frac{1}{2}(c^{\dag}_{2} - c_{2}) + \frac{1}{2}(c^{\dag}_{3} + c_{3})\\
& E_{3} = 4: \hspace{5mm}\Phi_{3} = \frac{1}{2}(c^{\dag}_{3} - c_{3}) + \frac{1}{2}(c^{\dag}_{4} + c_{4})\\
& E_{4} = 4: \hspace{5mm}\Phi_{4} = \frac{1}{2}(c^{\dag}_{4} - c_{4}) + \frac{1}{2}(c^{\dag}_{5} + c_{5})\\
& E_{5} = 0: \hspace{5mm}\Phi_{5} = \frac{1}{\sqrt{2}}(c^{\dag}_{1} + c_{1}) \\
& E_{6} = -8: \hspace{5mm}\Phi_{6} = \frac{1}{\sqrt{2}}c_{3} -\frac{1}{2\sqrt{2}}(c^{\dag}_{2} - c_{2}) + \frac{1}{2\sqrt{2}}(c^{\dag}_{4} + c_{4}) \\
& E_{7} = -8: \hspace{5mm}\Phi_{7} = -\frac{1}{2}(c^{\dag}_{2} - c_{2}) + \frac{1}{2}(c^{\dag}_{3} + c_{3})\\
& E_{8} = -4: \hspace{5mm}\Phi_{8} = -\frac{1}{2}(c^{\dag}_{3} - c_{3}) + \frac{1}{2}(c^{\dag}_{4} + c_{4}) \\
& E_{9} = -4: \hspace{5mm}\Phi_{9} =  -\frac{1}{2}(c^{\dag}_{4} - c_{4}) + \frac{1}{2}(c^{\dag}_{5} + c_{5})\\
& E_{10} = 0: \hspace{5mm}\Phi_{10} =  \frac{1}{\sqrt{2}}(c^{\dag}_{5} - c_{5}) \\
\end{split}
\end{equation*}

We analyze the emergence of Nambu fermion operator pairs in the mirror model. The energies occupying the bulk modes form pairs which are Hermitian conjugate to one another: $\Phi_{2}$ and $\Phi_{7}$, $\Phi_{3}$ and $\Phi_{8}$, $\Phi_{4}$ and $\Phi_{9}$. However, we observe the presence of unpaired operators: $\Phi_{5}$ and $\Phi_{10}$ which directly correspond to the known unpaired Majorana modes localized at zero energies which live at the edges of the spin chain \cite{Kitaev_2001}, and $\Phi_{1}$ and $\Phi_{6}$, which we identify as forming a topological mirror point localized on the third qubit. One can explicitly check that the Hermitian conjugate of $\Phi_{1}$ or $\Phi_{6}$ differs from each other by the fermionic creation/annihilation operator $c^{\dag}_{3}$/$c_{3}$ occupying site 3 of the lattice. 

The structure of the above spectrum is directly visible in the dynamics of the model. From our analysis, we find three localized states: two at zero energy, and conversely, one far from zero energy occupying E = $|8|$ which can be viewed as a topological mirror point. As expected, the two unpaired zero energy states which live on weakly linked edges are highly robust during open system time evolution due to protection via high degree of localization within the energy gap induced by staggered mirror couplings. We remark that when the quench is performed, these localized states will be shifted to a finite value instead of exactly zero. Conversely, the topological mirror point, which exists at energies far from zero, behaves as a bulk state experiencing minimal protection with weak temporal oscillations of the magnetization. These predictions are consistent with the closed system result for the mirror model obtained in Fig. \ref{fig:closed_2442}.

\subsection{Topological Defect Model}\label{sec:defect}
We design a topological defect model $\mathcal{H}_{D}$, depicted by Fig. 1(c). Similarly, we define the defect model $\mathcal{H}_{D}$ as a topologically non-trivial limit of Eq. \ref{eq:spin-chain} with specified couplings $J^{z}_{1} = 4$, $J^{z}_{2} = 2$, $J^{z}_{3} = 2$, $J^{z}_{4} = 4$, $J^{z}_{5} = 0$, with $h^{x} = 1$. We exactly follow the same procedure previously outlined in Sec. \ref{sec:mirror} to analyze the corresponding spectrum of $\mathcal{H}_{D}$. We solve the following BdG Hamiltonian obtained after performing the Jordan-Wigner transformation \ref{eq:JW} of $\mathcal{H}_{D}$  with OBC before the quench is performed $(t > 0)$,
\begin{equation}
    \langle\mathcal{H}_{D}^f \rangle= \Psi^{\dag} H \Psi = \Psi^{\dag} \begin{pmatrix}
    A & B &\\
    -B^{*} & -A^{*} &
     \end{pmatrix} \Psi \label{eq:BdG_D}
\end{equation}
where $H$ is the $2N \times 2N$ BdG block matrix containing matrices $A$ and $B$ satisfying $A = A^{\dag}$ and $B^{T} = -B$. We obtain the corresponding matrix $H$ for the defect model
\begin{equation}
H = \frac{1}{2}\begin{pmatrix}
  \begin{matrix}
   & -J' & & &  \\
   -J' & 0 & -J & &  \\
   & -J & 0 & -J&  \\
   &  & -J & 0& -J'  \\
   &  &  & -J' & 0 \\
  \end{matrix}
  & \rvline & \begin{matrix}
    & -J' &  & &  \\
   J' & 0 & -J & &  \\
   & J & 0 & -J &  \\
   &  & J & 0& -J'  \\
   &  &  & J' & 0 \\
  \end{matrix} \\
\hline
  \begin{matrix} 
  & J' &  & &  \\
   -J' & 0 & J & &  \\
   & -J & 0 & J&  \\
   &  & -J & 0& J'  \\
   &  &  & -J' & 0 \\
  \end{matrix} & \rvline &
  \begin{matrix}
  & J' &  & &  \\
   J' & 0 & J & &  \\
   & J & 0 & J &  \\
   &  & J & 0& J'  \\
   &  &  & J' & 0 \\
  \end{matrix}
\end{pmatrix}
\end{equation}

Following the earlier procedure, the eigenvectors corresponding to the solution of Eq. \ref{eq:BdG_D} can be organized into the unitary block diagonal matrix form $\mathcal{U}_{D}$ and can be analyzed by defining new Nambu fermion operators $\Phi = \mathcal{U}_{D} \Psi$ with the exact same structure as above. We obtain the following spectrum written in terms of the Nambu fermion operators:

\begin{equation*}
\begin{split}
\centering
& E_{1} = 8: \hspace{5mm}\Phi_{4} = \frac{1}{2}(c^{\dag}_{1} - c_{1}) + \frac{1}{2}(c^{\dag}_{2} + c_{2})\\
& E_{2} = 8: \hspace{5mm}\Phi_{2} = \frac{1}{2}(c^{\dag}_{4} - c_{4}) + \frac{1}{2}(c^{\dag}_{5} + c_{5})\\
& E_{3} = 4: \hspace{5mm}\Phi_{3} = \frac{1}{2}(c^{\dag}_{2} - c_{2}) + \frac{1}{2}(c^{\dag}_{3} + c_{3})\\
& E_{4} = 4: \hspace{5mm}\Phi_{1} = -\frac{1}{\sqrt{2}}c_{3} -\frac{1}{2\sqrt{2}}(c^{\dag}_{2} - c_{2}) + \frac{1}{2\sqrt{2}}(c^{\dag}_{4} + c_{4}) \\
& E_{5} = 0: \hspace{5mm}\Phi_{5} = \frac{1}{\sqrt{2}}(c^{\dag}_{1} + c_{1}) \\
& E_{6} = -8: \hspace{5mm}\Phi_{9} =  -\frac{1}{2}(c^{\dag}_{1} - c_{1}) + \frac{1}{2}(c^{\dag}_{2} + c_{2})\\
& E_{7} = -8: \hspace{5mm}\Phi_{7} = -\frac{1}{2}(c^{\dag}_{4} - c_{4}) + \frac{1}{2}(c^{\dag}_{5} + c_{5})\\
& E_{8} = -4: \hspace{5mm}\Phi_{8} = -\frac{1}{2}(c^{\dag}_{2} - c_{2}) + \frac{1}{2}(c^{\dag}_{3} + c_{3}) \\
& E_{9} = -4: \hspace{5mm}\Phi_{6} = \frac{1}{\sqrt{2}}c_{3} -\frac{1}{2\sqrt{2}}(c^{\dag}_{2} - c_{2}) + \frac{1}{2\sqrt{2}}(c^{\dag}_{4} + c_{4}) \\
& E_{10} = 0: \hspace{5mm}\Phi_{10} =  \frac{1}{\sqrt{2}}(c^{\dag}_{5} - c_{5}) \\
\end{split}
\end{equation*}

We identify key features of this spectrum which underlie the observed dynamics. First, in agreement with Sec. \ref{sec:mirror}, we find the emergence of Nambu fermion pairs which classify the bulk modes $\Phi_{1}$ and $\Phi_{6}$, $\Phi_{2}$ and $\Phi_{7}$, $\Phi_{3}$ and $\Phi_{8}$ which are Hermitian conjugates of themselves. We find unpaired Nambu fermion operators $\Phi_{5}$ and $\Phi_{10}$ which correspond to the unpaired Majorana modes found in the above models of Sec. \ref{sec:staggered} and Sec. \ref{sec:mirror}. Additionally, we find the Nambu fermion pairing $\Phi_{4}$ and $\Phi_{9}$ which are not Hermitian conjugate due to the creation/annihilation operator $c^{\dag}_{3}/c_{3}$ acting on site three, which we identify in this case to correspond to the topological defect on qubit 3 in the center of the chain. Therefore, we find the existence three topological modes in this model: two on the ends of the chain, and one defect mode in the center. 

Notably, one can characterize the difference in dynamics produced by the topological modes of $\mathcal{H}_{D}$ when compared to the other two models discussed in Sec. \ref{sec:staggered} and \ref{sec:mirror}. In our case, the topological defect model can be  viewed as a domain wall between two topological dimer configurations which introduces a pair of localized states at both ends of the chain, which in the 5-qubit case results in the producing of three distinct surface modes \cite{Estarellas_2017, PhysRevB.84.195452}. In contrast to the other two models, whose topological edge modes live on the weakly-coupled links, the two edge modes of the defect model live on strongly coupled links and hence experience a smaller degree of localization which is reflected in their open system dynamics. For the topological defect mode created by the union of two weakly-coupled links at the center of the model, we expect to observe signatures of topological localization due to their occupancy at energies nearer to zero, in contrast to the topological mirror point discussed in \ref{sec:mirror} which exists farther from zero energy and therefore no observable Fourier signature. This is accurately predicted by the closed system result in Fig. \ref{fig:closed_4224}.

\section{Constant-Depth Circuits Construction}\label{sec:cd_circuits_const}
\begin{figure} [hbt!]
\centering
    \includegraphics[scale = .22]{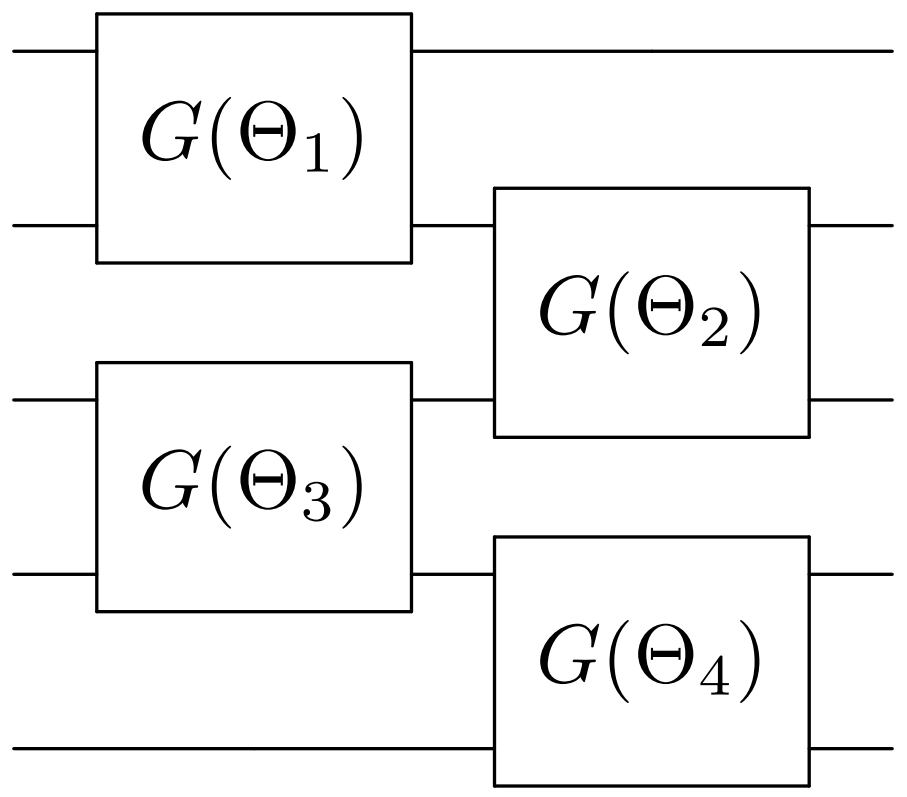}
    \caption{Parameterized circuit structure for a $5$-qubit system for $1$ time-step.}
    \label{fig:time_step1_basic}
\end{figure}

We construct constant depth circuits \cite{Bassman_Oftelie_2022} in the following manner. Since the simulated Hamiltonians are quadratic containing nearest neighbor interactions, they can be implemented in the form of two qubit gates acting on adjacent qubits. Using these gates, we construct the circuit for the time evolution of one time-step. As an example, for a $5$-qubit system the circuit is shown in Fig. \ref{fig:time_step1_basic}. 

Here, each $G(\Theta_i)$ is a two qubit gate parameterized by a set of parameters $\Theta_i$. Note that this parameter set is different for each gate $G$. Below, we will drop $\Theta_i$ in the labelling of the gates for convenience. 

For every subsequent time-step, the circuit in Fig. \ref{fig:time_step1_basic} needs to be repeated. It is straightforward to observe that for a time evolution of $n$ time-steps, there are $2n$ `columns' of the $G$ gates. To downfold the circuit, i.e. to reduce the number of columns of the $G$ gates, thereby reducing the circuit depth, we can exploit special properties of these gates. 
The $G$ gates belong to a special group of two qubit gates known as matchgates \cite{valiant_matchgate}, which we discuss below. We report here only salient features that are needed for this work. 

\begin{definition}[Matchgate]
    Consider two matrices $A, B$ in $\mathit{SU}(2)$:
    \begin{align}
    A=\begin{bmatrix}
        a & b \\
        c & d
    \end{bmatrix}, \,\, B=\begin{bmatrix}
        e & f \\
        g & h
    \end{bmatrix},
    \end{align}
    where det($A$)=det($B$). Then the two qubit matchgate $G(A,B)$ is defined as 
    \begin{align}
    G = \begin{bmatrix}
        a & & & b \\
         & e & f & \\
         & g & h &  \\
        c & & & d 
    \end{bmatrix}    
    \end{align}
\end{definition}

\begin{figure} [hbt!]
\centering
    \includegraphics[scale = .22]{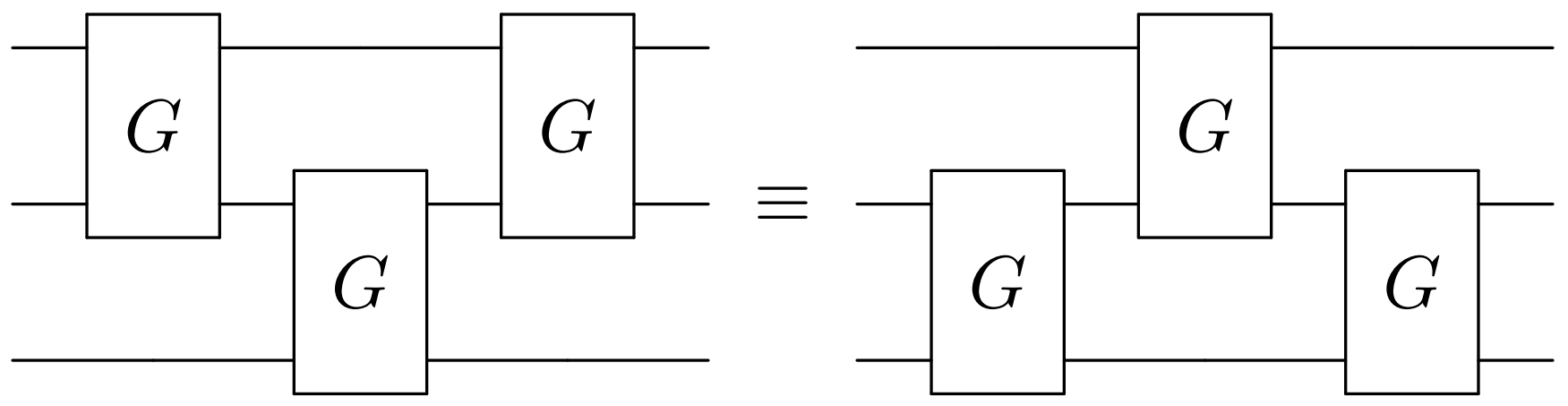}
    \caption{Illustration of the empirical conjecture $(G_{1} \otimes I)(I \otimes G_{2})(G_{3} \otimes I) = (I \otimes G_{4})(G_{4} \otimes I)(I \otimes G_{6})$ \cite{Bassman_Oftelie_2022}.}
    \label{fig:matchgate_order}
\end{figure}

Matchgates have two properties that allow for the downfolding of circuits. First, the product of two matchgates is a matchgate. The proof for this is obtained via matrix multiplication. The second property is an empirical conjecture that is found to be satisfied using numerical optimization for quadratic Hamiltonians. Namely, for distinct matchgates $G_{1}$, $G_{2}$, and $G_{3}$ there exist matchgates $G_{4}$, $G_{5}$, and $G_{6}$ of the same type such that $(G_{1} \otimes I)(I \otimes G_{2})(G_{3} \otimes I) = (I \otimes G_{4})(G_{4} \otimes I)(I \otimes G_{6})$ \cite{Bassman_Oftelie_2022}. This property is visualized with the circuit configuration shown in Fig. \ref{fig:matchgate_order}. 

\begin{figure} [hbt!]
\centering
    \includegraphics[scale = .22]{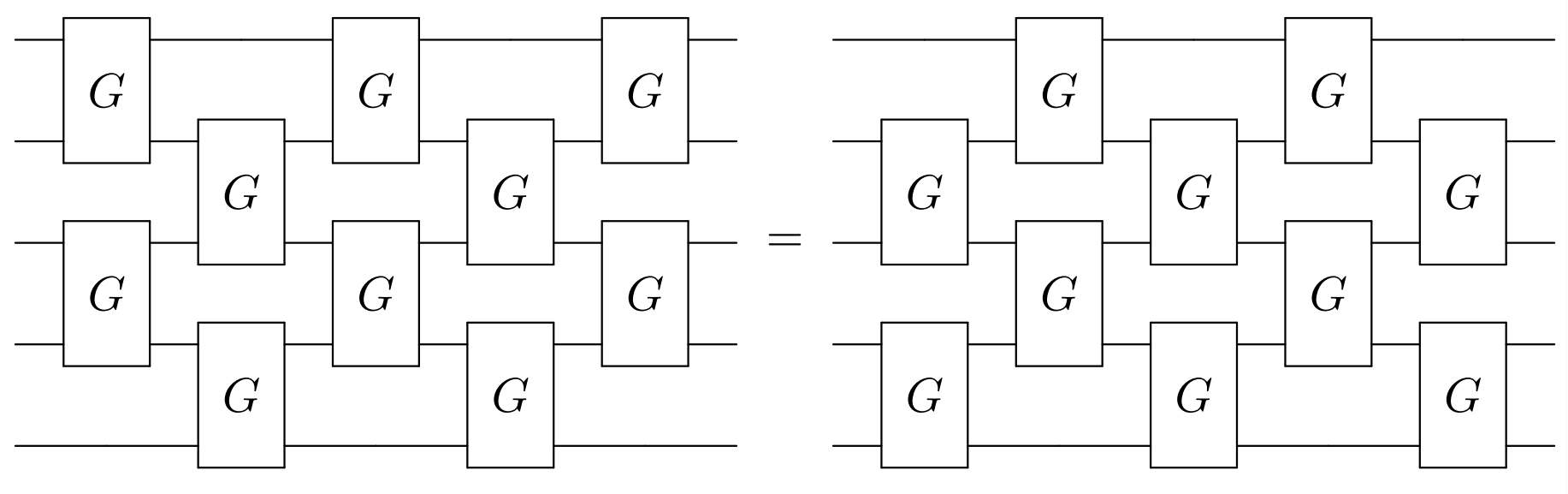}
    \caption{Matchgate mirror identity for $5$ qubits.}
    \label{fig:matchgate_mirror5}
\end{figure}

Using this conjecture we can construct mirror identities for matchgates. An example is presented in Fig. \ref{fig:matchgate_mirror5} for a $5$-qubit system, which is implemented to generate matchgates for the topological models depicted in Sec. \ref{sec:topo_models}.

\begin{figure}
    \centering
    \begin{subfigure}[b]{0.5\textwidth}
        \centering
        \includegraphics[scale=0.22]{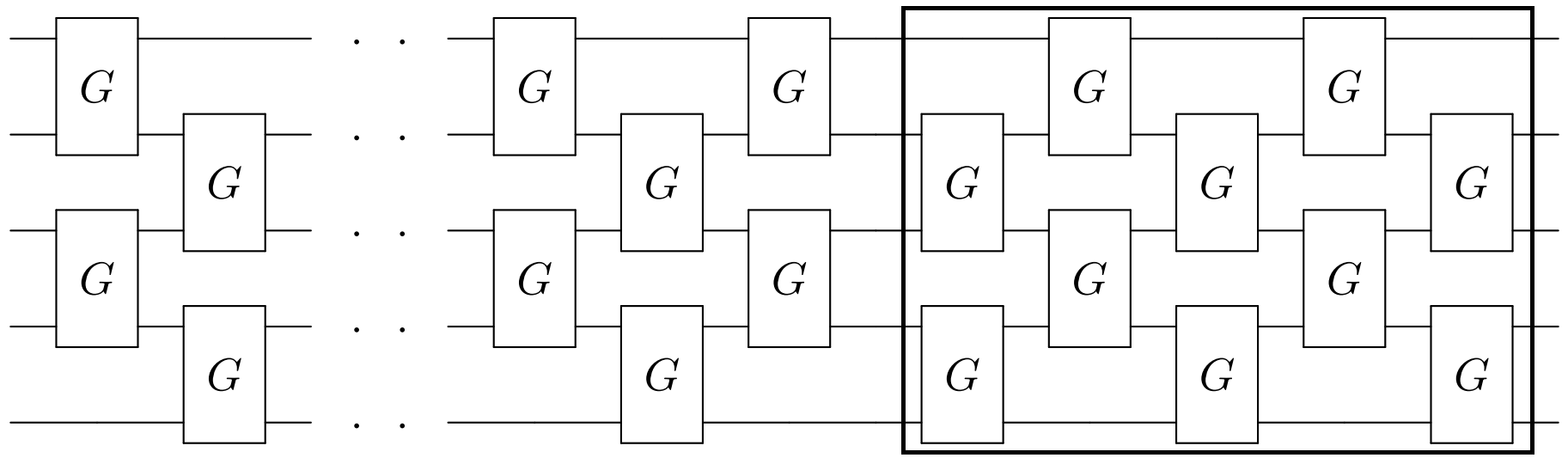}
        \caption{The $5$-qubit parameterized circuit for simulating $n$ time-steps. Note that it will contain $2n$ columns of matchgates.}
        \hspace{10cm}
        \label{fig:cd_comp_step1}
    \end{subfigure}
    \hfill
    
    \begin{subfigure}[b]{0.5\textwidth}
        \centering
        \includegraphics[scale=0.22]{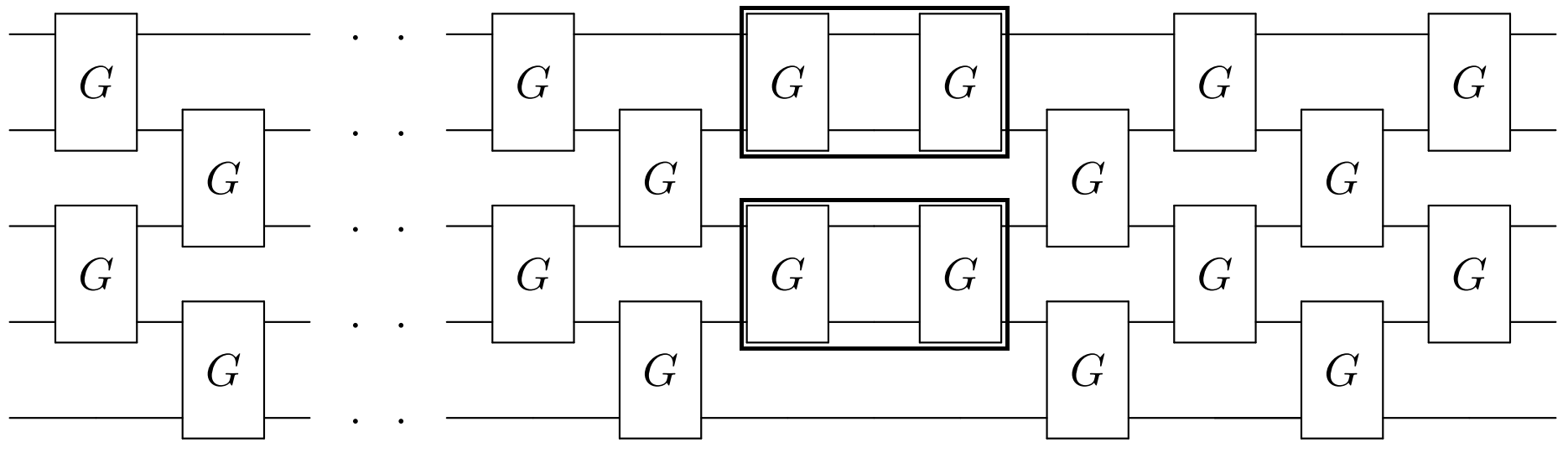}
        \caption{The circuit after applying the mirror identity to the last $5$ columns of the circuit in (a). The two sets of highlighted matchgates can be combined into one.}
        \hspace{10cm}
        \label{fig:cd_comp_step2}
    \end{subfigure}
    \hfill
    \begin{subfigure}[b]{0.5\textwidth}
        \centering
        \includegraphics[scale=0.22]{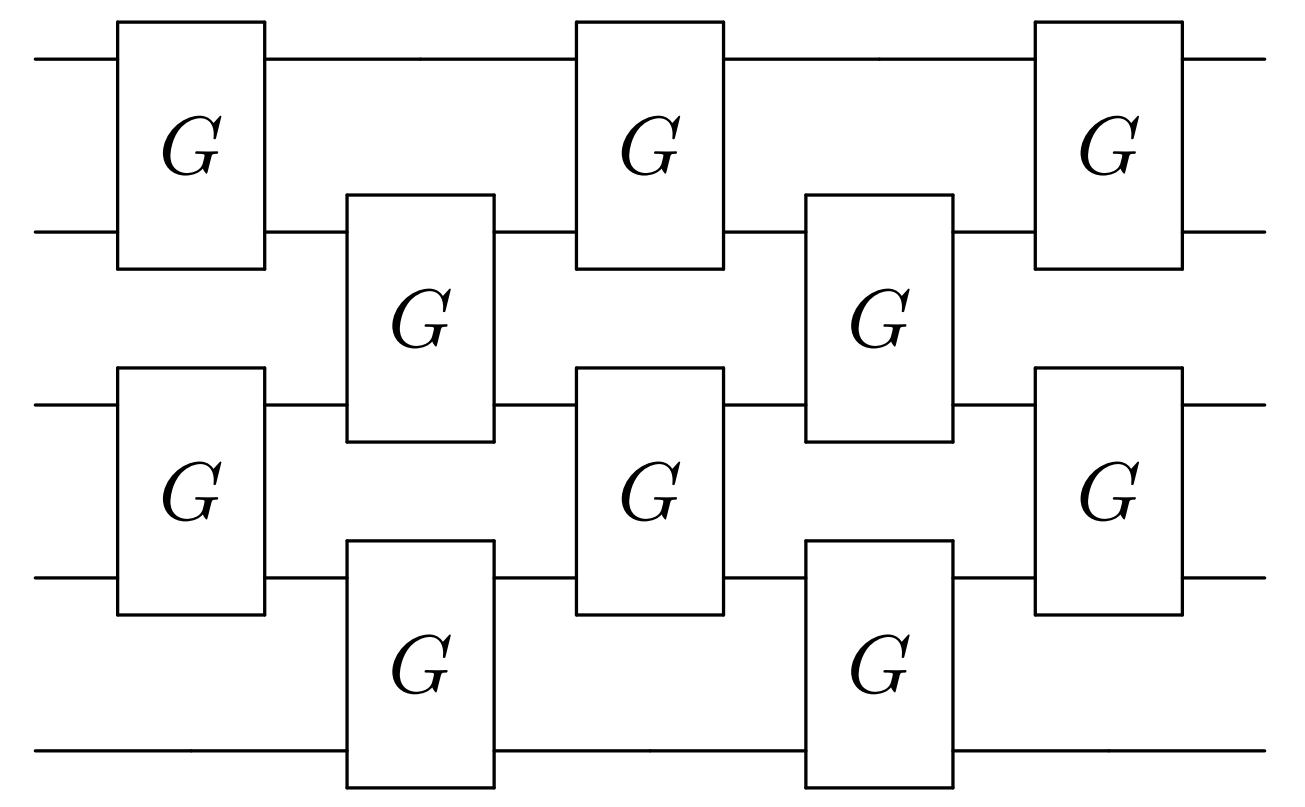}
        \caption{Final circuit after repeated application of the process in (a) and (b) until only $5$ columns of matchgates remain.}
        \label{fig:cd_comp_step3}
    \end{subfigure}
    \hfill
    \caption{Construction of constant depth circuits using matchgate properties for a $5$-qubit system. The Hamiltonian is simulated for $n$ time-steps.}
    \label{fig:cd_creation}
\end{figure}

To construct the constant depth circuits, we first create the $N$ qubit parameterized circuit for simulating $n$ time-steps that will contain $2n$ columns of the $G$ gates. Next we apply the mirror identity to the last $N$ columns of the $G$ gates. This allows the combination of two matchgates into one, and by extension allows the combination of two columns of matchgates into one. This process is repeated until $N$ columns of matchgates remain. This process for a $5$-qubit system is depicted in Fig. \ref{fig:cd_creation}. Note that irrespective of $n$, the final circuit will have $N$ columns of matchgates, leading to a circuit depth scaling that is constant with respect to the number of time steps. 

In practice, we directly construct the parameterized circuit similar to the one in  Fig. \ref{fig:cd_comp_step3}. As mentioned previously, there are $N$ columns of matchgates for the $N$ qubit system. Next, we compute the equivalent matrix of this parameterized circuit which is then compared with the desired unitary evolution operator to find the optimized circuit parameters. The desired unitary evolution operator is obtained by exact diagonalization. Note that every time-step has a separate constant depth circuit. For example, if a system has to be simulated according to a particular Hamiltonian for $100$ time-steps, then there are $100$ constant depth circuits. These circuits are independent, and hence their optimization can be executed in parallel. 

\begin{figure} [hbt!]
\centering
    \includegraphics[scale = .4]{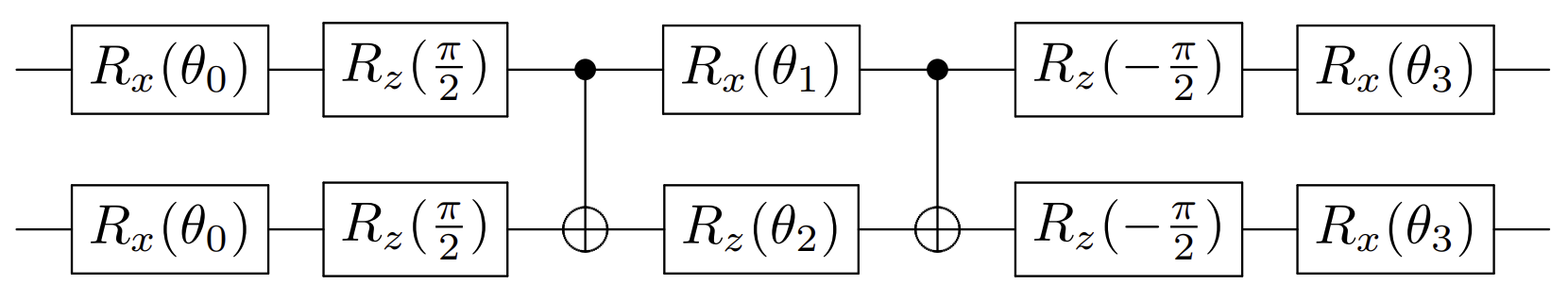}
    \caption{Decomposition of the matchgate corresponding to the Hamiltonian simulated in our model.}
    \label{fig:matchgate_tfim}
\end{figure}

The matchgates themselves are decomposed into native gate circuits. For certain Hamiltonians \cite{Bassman_2021}, these matchgates require only two CNOT gates, unlike a general two qubit gate that requires at most three CNOT gates. For the $N$ qubit system, the constant depth circuit contains $N(N-1)$ CNOT gates. Decompositions of matchgates into the native gates for different Hamiltonians are provided in \cite{Bassman_Oftelie_2022}. Fig. \ref{fig:matchgate_tfim} shows the decomposition for the matchgates corresponding to the Hamiltonian the we have studied in this work.
Each $\theta_i$ set is determined after optimizing the constant depth circuits for the Hamiltonian of interest. $R_x(\theta), R_z(\theta)$ are rotation gates that rotate the qubit about the $x, z$ axis respectively by angle $\theta$. Further details on the construction of the optimized circuits are given in Sec. \ref{sec:qhardware}. 

\begin{figure} [hbt!]
\centering
    \includegraphics[scale = .4]{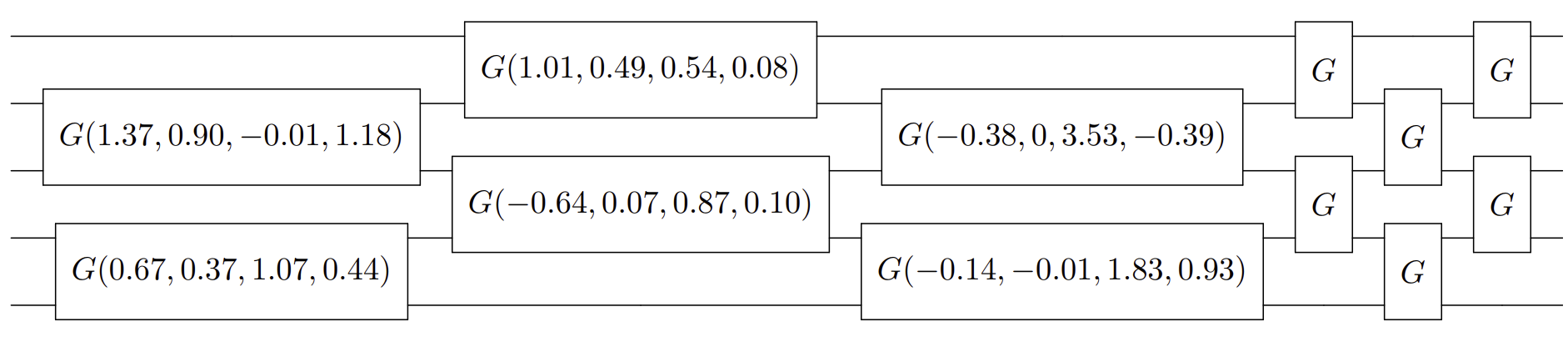}
    \caption{Optimized constant depth circuit for a $5$ qubit system for the simulation upto the first time-step. Each $G$ gate is of the form in Fig. \ref{fig:matchgate_tfim}. Note that these optimized parameters correspond to a subset of the constructed total gates.}
    \label{fig:matchgate_mirror5_opt_param}
\end{figure}

It is observed that adding an additional column of matchgates leads to better results. Hence our results with the constant depth circuits have $N+1$ columns of matchgates for a $N$ qubit system instead of $N$.
An example of an optimized circuit is shown in Fig. \ref{fig:matchgate_mirror5_opt_param}. Here, the circuit is for the first time-step for a $5$-qubit system. Each $G$ is of the form in Fig. \ref{fig:matchgate_tfim} with $4$ optimized parameters. We show a few optimized parameters as an example, noting that this optimization can be performed independent of the algorithm employed. We clarify that the ability to numerically optimize constant-depth circuits for a given spin chain using this particular unitary matrix fitting method is generally dependent on the number of qubits in the system. However, by employing more sophisticated classical optimization algorithms, one can address potential size-dependent computational bottlenecks and fit circuits for feasibly larger systems within reasonable resource cost \cite{Bassman_Oftelie_2022}.

\section{Quantum Hardware Implementation}\label{sec:qhardware}

\subsection{Methodology}
We implement the 1D Quantum Spin Hamiltonian given in Eq. 1 in the main text for a $5$-qubit system for different values of $J, J'$ . We consider a total number of $100$ time-steps. Each time-step has a value of $0.1$ units of inverse magnetic field $\hbar/h_{x}$ following the convention $\hbar=1$. Note that each time-step has a separate parameterized constant depth circuit leading to a total of $100$ parameterized constant depth circuits that are optimized as described in Sec. \ref{sec:cd_circuits_const}. 

We begin by creating $100$ target matrices for the $100$ constant depth circuits using exact diagonalization. Then, we create $100$ parameterized constant depth circuits. The optimal parameters for each of these $100$ circuits are calculated by maximizing the `overlap' (i.e. minimizing a distance function) between the target matrix and the equivalent matrix of the constant depth circuit. This distance function, motivated by the definition of the Hilbert-Schmidt inner product, can be defined as follows. Letting 
$d_{U}$ be the dimension of the (unitary) target matrix $U_{T}$ and parameterized constant-depth circuit matrix $U$, their distance is given by 

\begin{equation}
    D(U, U_{T}) = 1 - \frac{Tr(U^{\dag}U_{T})}{d_{U}}
    \label{eq:distance_matcgates}
\end{equation}

The closer $U^{\dag}U_{T}$ is to the identity $I_{d_{U}}$, $U^{\dag}U_{T} \approx I_{d_{U}}$, the more indistinguishable the parameterized matrix $U$ is from the target matrix $U_{T}$ is. When $U^{\dag}U_{T} = I_{d_{U}} $, $D(U, U_{T}) = 0$ as expected, and hence the parameterization is exact. For technical details the reader is referred elsewhere \cite{Davis2020TowardsOT}. Once the optimal parameters are calculated, the final optimal constant depth circuit for each of the time-step is constructed. We point out that all of these tasks along with the execution of the circuits on IBM Quantum hardware can be performed directly using Python and the IBM \emph{Qiskit} software package \cite{Qiskit}. 
To summarize our methodology, we identify five steps in the computational protocol to follow the Hamiltonian dynamics of our chosen spin chain on the IBM quantum hardware, Fig. \ref{fig:hardware_flow}. 

\begin{figure}[htbp]
    \centering

\begin{center}
\begin{tabular}{|l |} 
 \hline
 \centerline{\textbf{Implementing Constant-Depth Circuits on Quantum Hardware}}\\[1ex]
 \hline\hline
 1. Construct $n$ constant-depth circuits for $n$ desired time steps of the $\mathcal{H}$ evolution. \\ [1ex]
 \hline
 2. Create a target matrix $U_{T}$ for each constant-depth circuit. \\ [1ex]
 \hline
 3. Generate parameterized constant-depth circuits by minimizing a distance function (Eq. \ref{eq:distance_matcgates}) between the target matrix $U_{T}$ \\ and constant-depth circuit matrix $U$. \\ [1ex]
 \hline
 4. Transpile parameterized circuits for a given backend. \\ [1ex]
 \hline
 5. Execute circuits on backend.\\ [1ex] 
 \hline
\end{tabular}
\end{center}

\caption{Flow of our computational protocol for implementing constant-depth circuits on quantum hardware.} \label{fig:hardware_flow}
\end{figure}

We now turn to specific details regarding hardware implementation of our digital quantum simulation protocol. IBM quantum processors are based on transmon superconducting qubit architectures. Gate operations are executed by performing sequences of microwave pulses to control states of system qubits which are physically realized via nonlinear LC circuits with Josephson junctions \cite{roth2021introduction}. 

\begin{figure} [hbt!]
\centering
    \includegraphics[scale = .4]{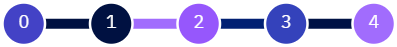}
    \caption{Qubit geometry of \textit{ibmq\_manila} \cite{Qiskit}, the quantum processor used in this work. The color code complies with IBM settings \cite{Qiskit}.}
    \label{fig:ibmq_manila_layout}
\end{figure}

We execute these circuits implementing our methodology by utilizing the IBM Quantum Processor \textit{ibmq\_manila}, which is accessible online to run experiments via the IBM Quantum Network \cite{Garc_a_P_rez_2020}. The reported average energy relaxation time of a qubit T1 for \textit{ibmq\_manila} is $100.56$$\mu s$, dephasing time T2 is $101.29$$\mu s$; the reported average readout fidelity is $0.9739$ and CNOT fidelity is $0.99$ \cite{Wang2022SoKBT}. The qubit layout for the device is shown in Fig. \ref{fig:ibmq_manila_layout}. We point out that while as of 2023 Sep 26th, \textit{ibmq\_manila} is retired and is no longer accessible via the IBM cloud, our results have been tested and shown to be consistent between different quantum processors and simulators provided by IBM. 

Following convention, we define 
$\ket{0} = \begin{bmatrix}
           1 \\
           0
\end{bmatrix} $
and 
$\ket{1} = \begin{bmatrix}
           0 \\
           1
\end{bmatrix} $. 
In each of our simulations, for simplicity, we choose the initial state for the circuits to be $\ket{00000}$. Operationally, \emph{Qiskit} follows a flipped notation for qubits represented by state kets, i.e., the state of the first qubit is denoted by the right-most position in the ket and is read right to left. As an example, for the initial state consisting of three qubits, $\ket{001}$, the qubit 1 is in state $\ket{1}$, qubit 2 in state $\ket{0}$, and qubit 3 in state $\ket{0}$. 

We execute a simulation known as a quantum quench. We initiate the ground state of our Hamiltonian with the applied magnetic field turned off: $\mathcal{H}(t < 0) = - \sum_{i=1}^{N-1} J_{z, i} Z_{i} Z_{i+1}$. Then, at $t=0$, the field is instantaneously turned on, and the evolution of the final Hamiltonian  $\mathcal{H}(t \geq 0) = - \sum_{i=1}^{N-1} J_{z, i} Z_{i} Z_{i+1} -  h_{x}\sum_{i=1}^{N} X_{i}$ is executed. During evolution, series of projective measurements are performed on each time-step for the observable $\sigma^{z}_{i}$, for each qubit $i = \{1,2,3,4,5\}$. To ensure that measurement outcomes are sampled with statistical sufficiency, repetitions of each circuit, known as shots, are executed along with their corresponding measurements. We then define our observable of interest, the local magnetization in the $z$-direction at each site, given by the corresponding expectation value 

\[ \langle \sigma^{z}_{i} \rangle = p_{i}(0) - p_{i}(1)\]  

where $p_{i}(0)$ and $p_{i}(1)$ are the averaged probabilities of obtaining the result 0 and 1 respectively from performing a projective measurement on qubit $i$ over all shots.
In our simulations, a total of $8192$ shots are executed to obtain the average the magnetization for each time step. 

Fig. 3 of the main text reports the results obtained with this approach for the dynamics of the spin model illustrated in Fig. 1(b) of the main text and described in Sec. IB of this document. 
Figs. \ref{fig:(4242)_cd} and \ref{fig:(4224)_cd} display the results obtained for the dynamics of the spin models illustrated in Figs. 1(a) and 1(c) of the main text and described in Sec. IA and Sec. IC of this document, respectively. 
In Fig. \ref{fig:(2222)_cd}, we show results for a quantum spin chain with uniform coupling. As in Fig. 1 of the main text, in Figs. \ref{fig:(4242)_cd}, \ref{fig:(4224)_cd} and \ref{fig:(2222)_cd} the plots in the left column show the observed on-site magnetization as a function of time. Applicable equilibration time methods are discussed in Sec. \ref{sec:eq_time}. The plots in the right column show the corresponding Fourier-transformed power spectra of the on-site magnetization. 

\subsection{Additional Results}\label{sec:additional}
Implementing the procedure discussed above, we comment on the dynamics of open system evolution for the models depicted by Fig. 1(a)  and Fig. 1(c) (Sec. \ref{sec:staggered} and Sec. \ref{sec:defect} respectively) of the main text. 

\begin{figure} [hbt!]
\centering
    \includegraphics[scale = .44]{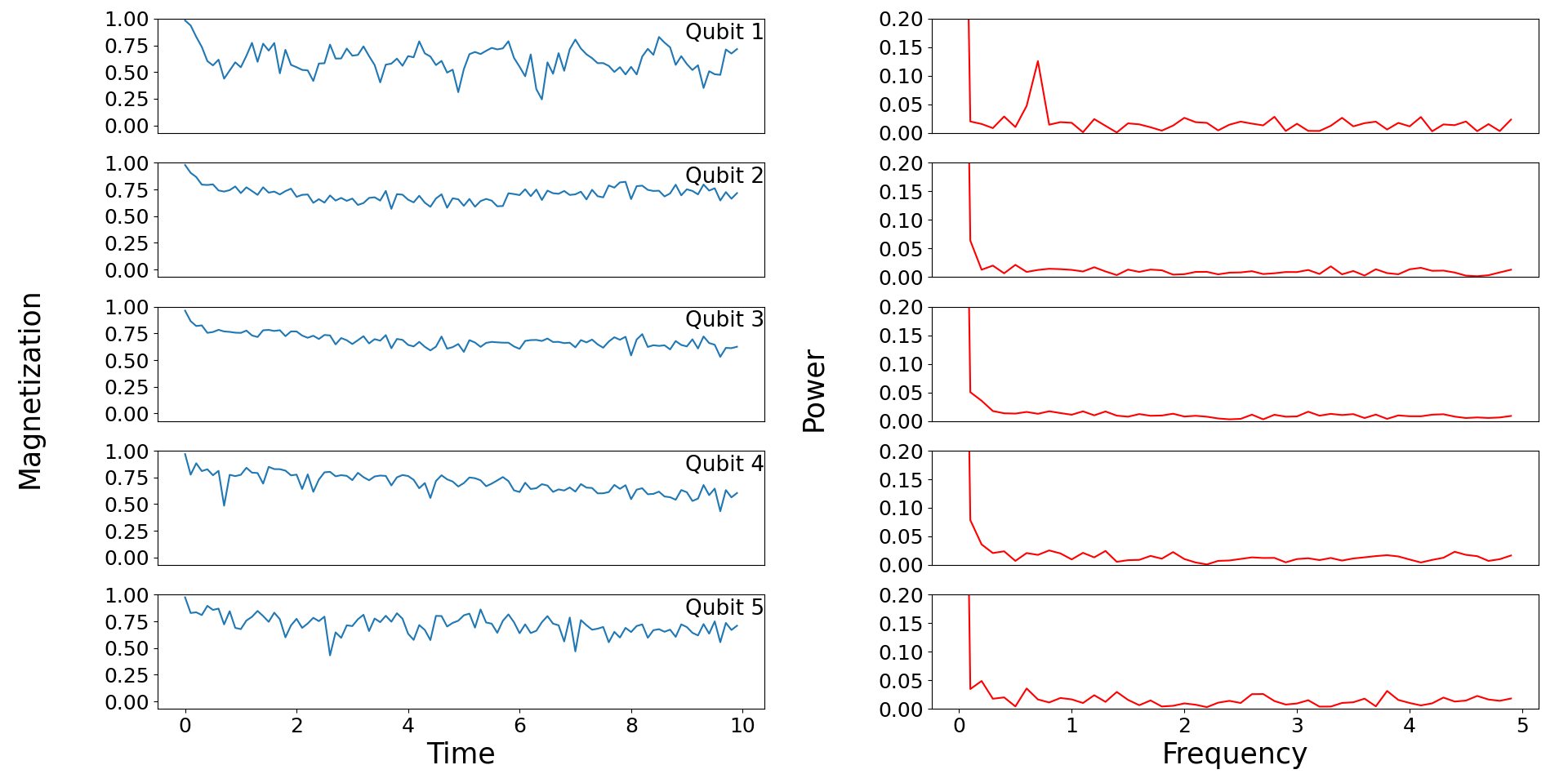}
    \caption{Local dynamics of a 5-qubit Hamiltonian in a topologically non-trivial regime, illustrated in Fig. 1(a) of the main text, obtained by NISQ processors. $J = 2$, $J' = 4$, $h_x = 1$. Left: real-space magnetization in the $z$-direction for each qubit. Time is reported in units of the inverse magnetic field $1/h^{x}$. Right: Fourier-transformed power spectrum in frequency space. A sharp surface mode signature is clearly visible on the first qubit, as expected based on the theory presented in Sec. \ref{sec:staggered}. Frequency is reported in units of the magnetic field $h^{x}$.} \label{fig:(4242)_cd}.
\end{figure}

The dynamics for the staggered coupling model are shown in Fig. \ref{fig:(4242)_cd}. In analogy with the mirror model shown as Fig. 3 in the main text, strong temporal oscillations of the magnetization of qubit 1 produce a strong peak in the frequency spectrum, denoting the presence of a topological mode localized to qubit 1. This is further supported by the observation that this peak occurs at comparable value of power and frequency as the two peaks observed in Fig. 3. 
We infer that constant-depth circuits have capability of stabilizing the topological mode, resulting in a long-time simulation where the robustness of qubit 1 is visible through transient dynamics. 

\begin{figure} [hbt!]
\centering
    \includegraphics[scale = .44]{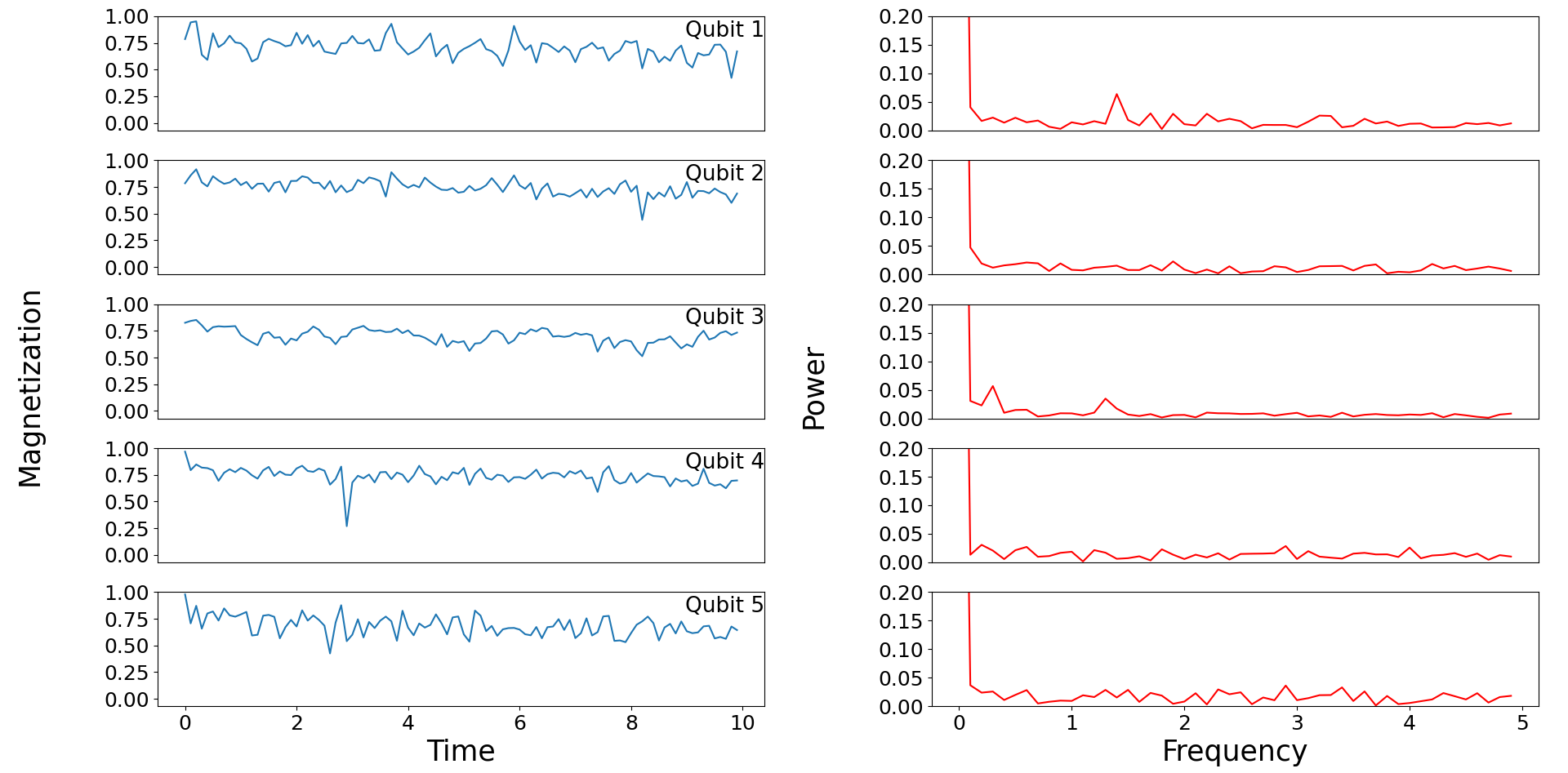} 
    \caption{Local dynamics of a 5-qubit Hamiltonian in a topologically non-trivial regime, illustrated in Fig. 1(c) of the main text, obtained by NISQ processors. $J = 2$, $J' = 4$, $h_x = 1$. Left: real-space magnetization in the $z$-direction for each qubit. Time is reported in units of the inverse magnetic field $1/h^{x}$. Right: Fourier-transformed power spectrum in frequency space. A topological signature imprinted by the defect mode is visible on the center qubit at frequency $\approx$ 1.2 units of magnetic field $h^{x}$. The bulk and boundary states delocalize due to strong outer couplings in conjunction with finite-size effects (compare to Figs. \ref{fig:closed_4224} and \ref{fig:closed_242242}).}
    \label{fig:(4224)_cd}
\end{figure}

The dynamics for the topological defect model are shown in Fig. \ref{fig:(4224)_cd}. The signature of a topological defect mode is visible on the center qubit indicated by the small peak lying in the interval of frequency $(1, 2)$, which is consistent with its corresponding closed system dynamics simulation result in Fig. \ref{fig:closed_4224}. We also identify a slight peak in the power spectrum of qubit 1 lying in a comparable frequency range. 
We observe that open system dynamics affects topologically non-trivial states hierarchically. As discussed in Sec. \ref{sec:defect}, the topological 'trimer' defect \cite{Estarellas_2017, PhysRevB.84.195452}, which can be viewed as a domain wall between two topologically distinct configurations, introduces a pair of localized states at both ends of the chain. While traces of the three localized states of qubits 1 and 3 are intelligible, noise limits any sharp values from being visible in the corresponding power spectrum, particularly on qubit 5. The degree of topological protection against noise is heavily limited in comparison to the models depicted in Fig. 1(a) and Fig. 1(b) of the main text, while remnants of protection are visible in comparison to the uniform case with no such features. Moreover, the bulk and boundary of the system hybridize as a result of open system dynamics. The finite size of the system has limited the degree of separation between the bulk and boundary of the system. This is corroborated by the findings in Fig. \ref{fig:closed_242242}, where the localized states of the edge have a larger degree of separation from the center of the system.

\begin{figure} [hbt!]
\centering
    \includegraphics[scale = .44]{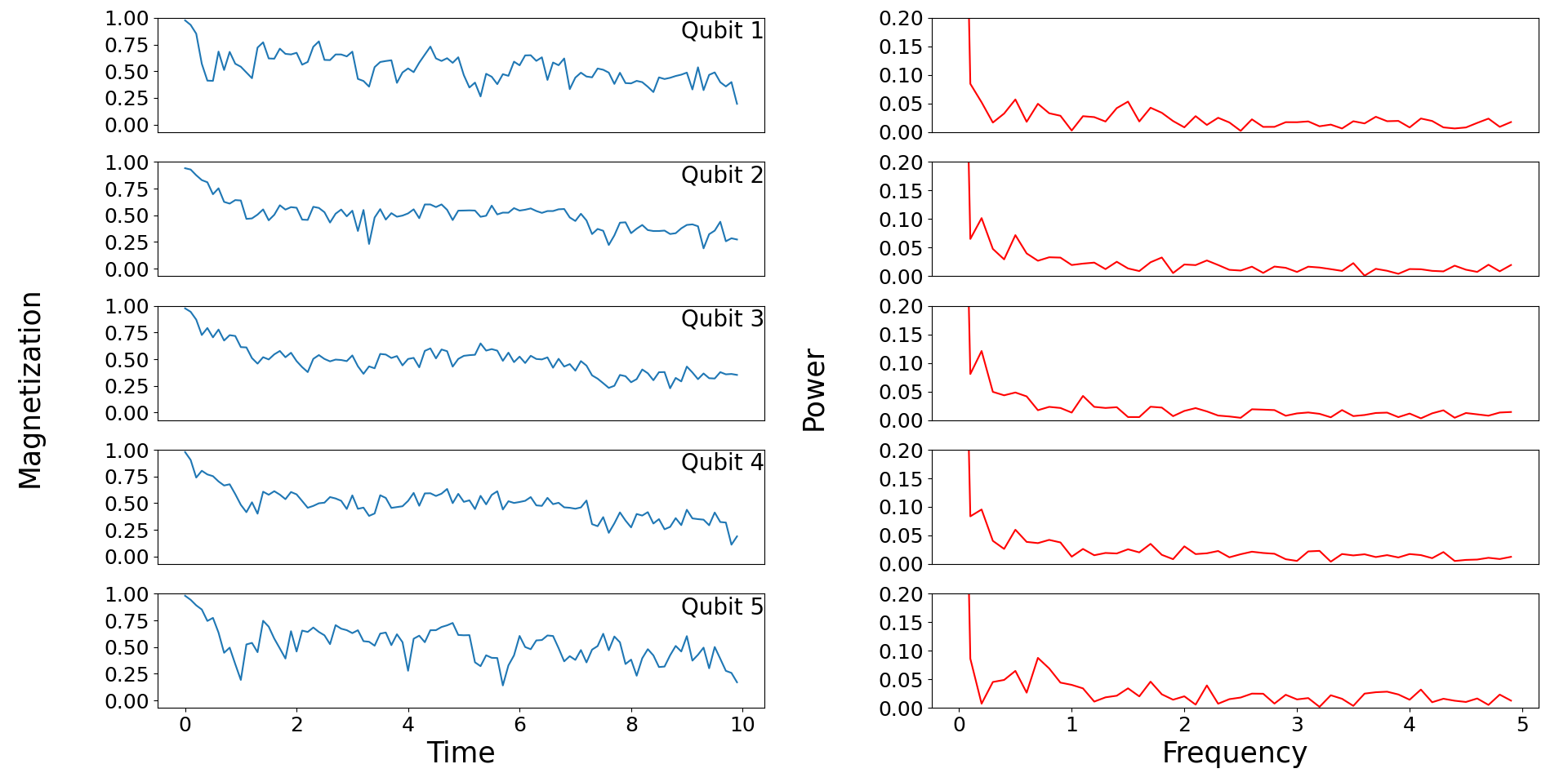}
    \caption{Local dynamics obtained from quantum hardware for a uniform coupling model, $J=J'=2$, $h=1$. This is equivalent to the non-topological regime of $\mathcal{H}$ for each model discussed in the main text. Time is reported in units of the inverse magnetic field $1/h^{x}$, frequency is reported in units of the magnetic field $h^{x}$. As expected, the absence of topological modes is evidenced by the noisy power spectrum.}
    \label{fig:(2222)_cd}
\end{figure}

Finally, we present as a benchmark in Fig. \ref{fig:(2222)_cd} the on-site magnetization in the time and frequency domains obtained from the dynamics simulation of a uniform 5-site spin chain. In this configuration, $J =J'$, and hence the absence of dimerization means that there are no topological features. As expected, 
the power spectra show that each qubit is indistinguishable from the rest of the system. We conclude by noting that, in each case, the observed large DC (zero energy) feature in the power spectrum is due to the polarization of the qubit spins along the $z$-direction, as pointed out in the main text.

\break

\subsection{Trotterization}\label{sec:trotter}

In this section, we present the results for the time-dependent on-site magnetization obtained with Hamiltonian dynamics simulations utilizing traditional Trotterization \cite{RevModPhys.86.153}. This method exploits the Suzuki-Trotter decomposition to break the total circuit implementation for the time evolution operator into discrete Trotter steps \cite{Smith_2019} (Eq. \ref{eq:trotter}). 

\begin{equation}
    e^{-i\mathcal{H}\Delta t} = \prod_{n} e^{-i\mathcal{H}_{n} \Delta t} + {\mathcal{O}(\frac{1}{M^{2}})}
    \label{eq:trotter}
\end{equation}

\begin{figure} [hbt!]
\centering
    \includegraphics[scale = .44]{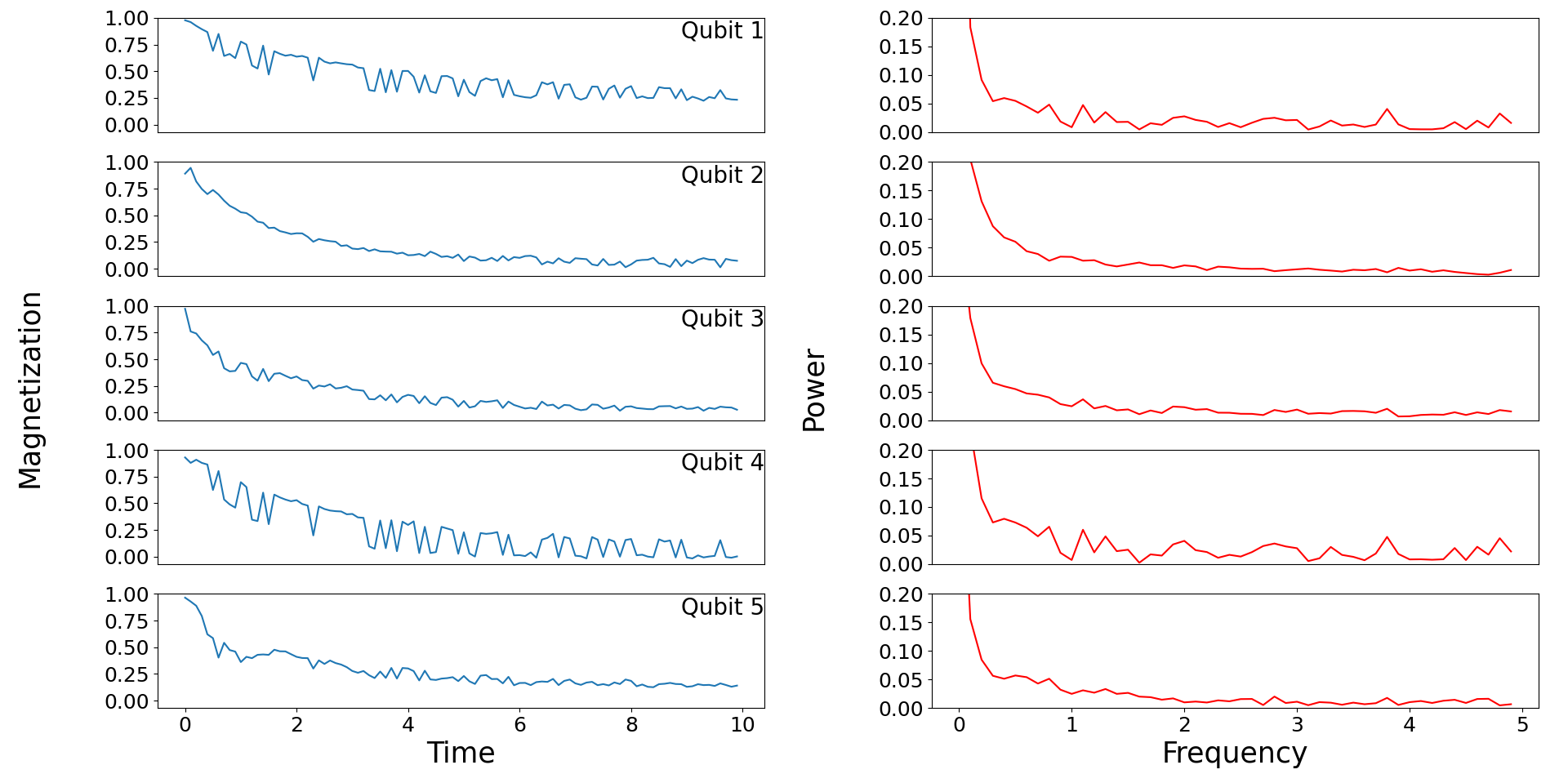}
    \caption{Local dynamics obtained from quantum hardware for the staggered spin model of Fig. 1(a), obtained with Trotterization. Time is reported in units of the inverse magnetic field $1/h^{x}$, frequency is reported in units of the magnetic field $h^{x}$. No clear frequency peak occurs in the power spectrum, at odds with the corresponding Fig. \ref{fig:(4242)_cd}.}
\label{fig: trotter_staggered}
\end{figure}

\begin{figure} [hbt!]
\centering
    \includegraphics[scale = .44]{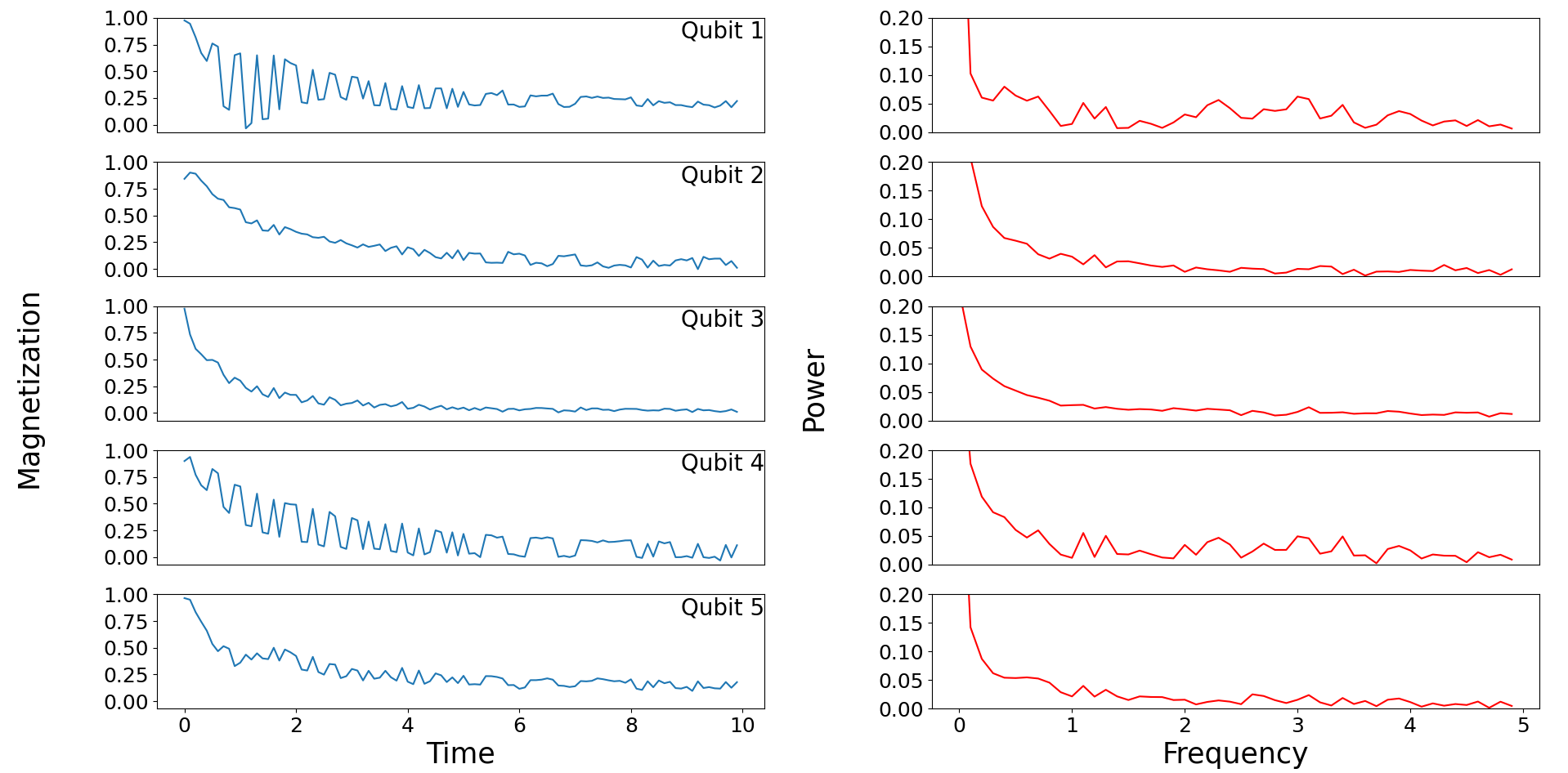}
    \caption{Local dynamics obtained from quantum hardware for the topological mirror spin model of Fig. 1(b), obtained with Trotterization. Time is reported in units of the inverse magnetic field $1/h^{x}$, frequency is reported in units of the magnetic field $h^{x}$. No clear frequency peak occurs in the power spectrum, at odds with the corresponding Fig. 3 of the main text.}
\label{fig: trotter_mirror}
\end{figure}

\begin{figure} [hbt!]
\centering
    \includegraphics[scale = .44]{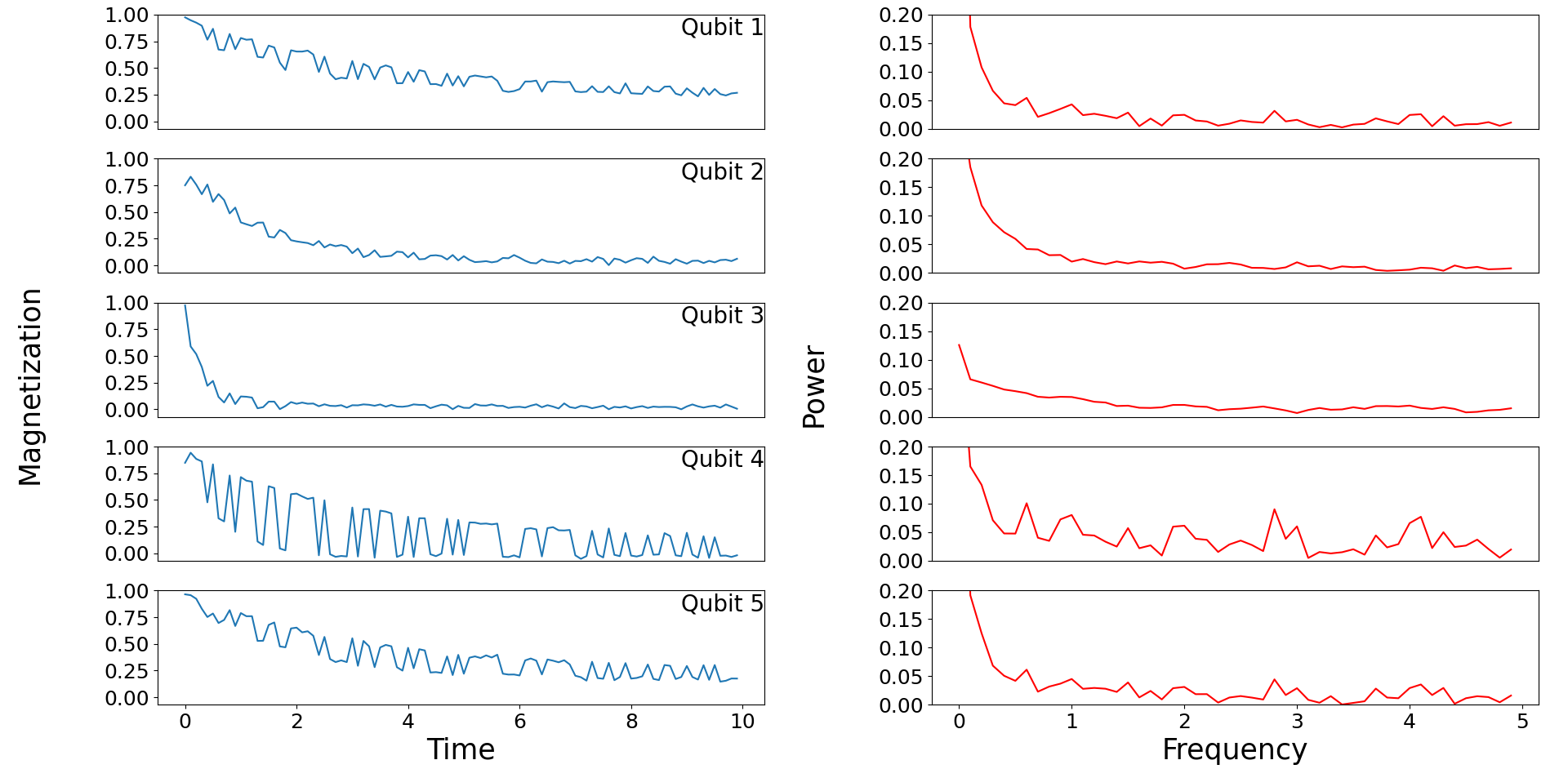}
    \caption{Local dynamics obtained from quantum hardware for the topological defect spin model of Fig. 1(c), obtained with Trotterization. Time is reported in units of the inverse magnetic field $1/h^{x}$, frequency is reported in units of the magnetic field $h^{x}$. No clear frequency peak occurs in the power spectrum, at odds with the corresponding Fig. \ref{fig:(4224)_cd}.}    
\label{fig: trotter_defect}
\end{figure}

\begin{figure} [hbt!]
\centering
    \includegraphics[scale = .44]{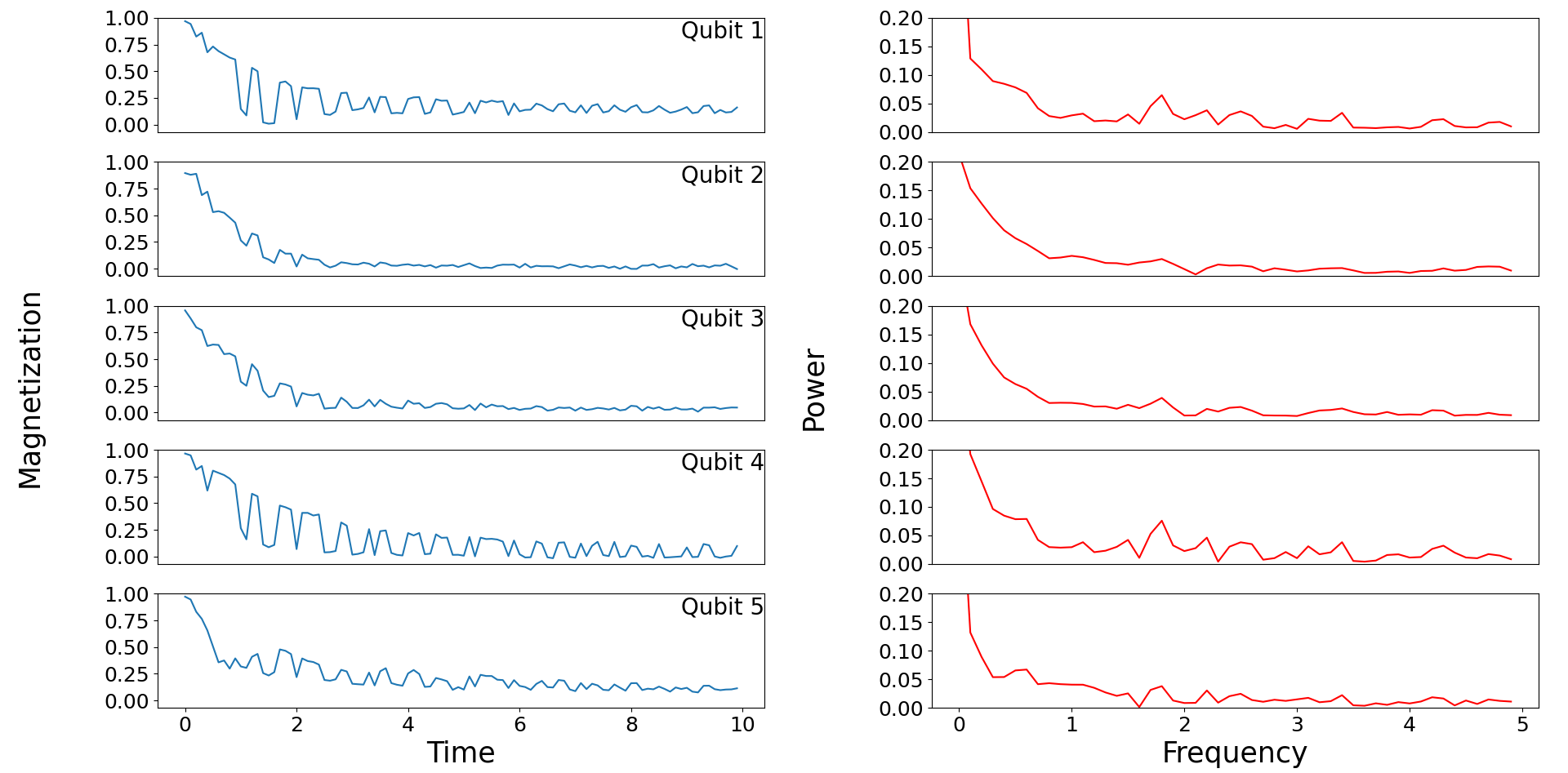}
    \caption{Local dynamics obtained from quantum hardware for for a uniform coupling model, obtained with Trotterization. Time is reported in units of the inverse magnetic field $1/h^{x}$, frequency is reported in units of the magnetic field $h^{x}$. No clear frequency peak occurs in the power spectrum, in line with Fig. \ref{fig:(2222)_cd}.}
\label{fig: trotter_constant}
\end{figure}

The time evolution of $\mathcal{H}$ is broken up into steps $e^{-i\mathcal{H}_{n} \Delta t}$ over which the instantaneous Hamiltonian $\mathcal{H}$ is constant. Each step is implemented directly into the circuit, resulting in an error, known as Trotter error, that scales with the number of time steps $M$ as $\mathcal{O}(\frac{1}{M})$. The corresponding dynamics obtained using Trotterization for each model in Fig. 1 of the main text and of the uniform quantum spin chain \ref{fig:(2222)_cd} are shown in Figs. \ref{fig: trotter_staggered}, \ref{fig: trotter_mirror}, \ref{fig: trotter_defect}, and \ref{fig: trotter_constant}, respectively. The individual time-step value used in the Trotterization simulations is chosen to be $0.1$ in units of inverse magnetic field $\hbar/h_{x}$, likewise following the convention $\hbar=1$.

In all cases, the circuits with time-dependent depth generate dynamical simulations which are inconsistent with the ground truth of each model that is presented in Sec. \ref{sec:closed_dynamics}. Using the downfolding property of constant-depth circuits, however, eliminates this issue 
and yields theory-consistent magnetization data. 
We can rephrase this evidence in terms of reduction of the Trotter error. In fact, the matchgates allow us to afford reasonably long simulations times even with tiny discretized time intervals. 
Compared to existing methods, our method produces viable results without the requirement of error mitigation \cite{Takagi_2022}. In other words, our method makes long-time dynamical simulations achievable intrinsically at the circuit-level.

\break
\subsection{Equilibration Time}\label{sec:eq_time}
The equilibration time $T_{eq}$ is the time required for the expectation value of an observable to approach its average, $\overline{\mathcal{A}}$, and fluctuate about it with fluctuations $\delta\mathcal{A} = \sqrt{\overline{\mathcal{A}^{2}} - \overline{\mathcal{A}}^{2}}$, which are due to finite system size \cite{equilibration_time}. Operationally, we calculate $T_{eq}$ by obtaining the first time when $\mathcal{A}(t) = \overline{\mathcal{A}}$. This choice is justified as a simple, intuitive measure for equilibriation times in gapped quantum systems comparable to ours, where constant-depth circuits generate  results where the oscillatory dynamics can stablize over a long period, and does not rapidly decay according to an exponential function. This definition is suitable for finite-size systems, noting that $T_{eq}$ scales with the system size $N$ when in a critical state, and size-independent when in a gapped or clustering initial state \cite{PhysRevA.87.032108}. We remark, however, that the equilibriation time is a measure with a degree of arbitrariness, whose usefulness depends directly on the system itself. 

\begin{figure}[hbt!]
\centering
    \includegraphics[scale = .4]{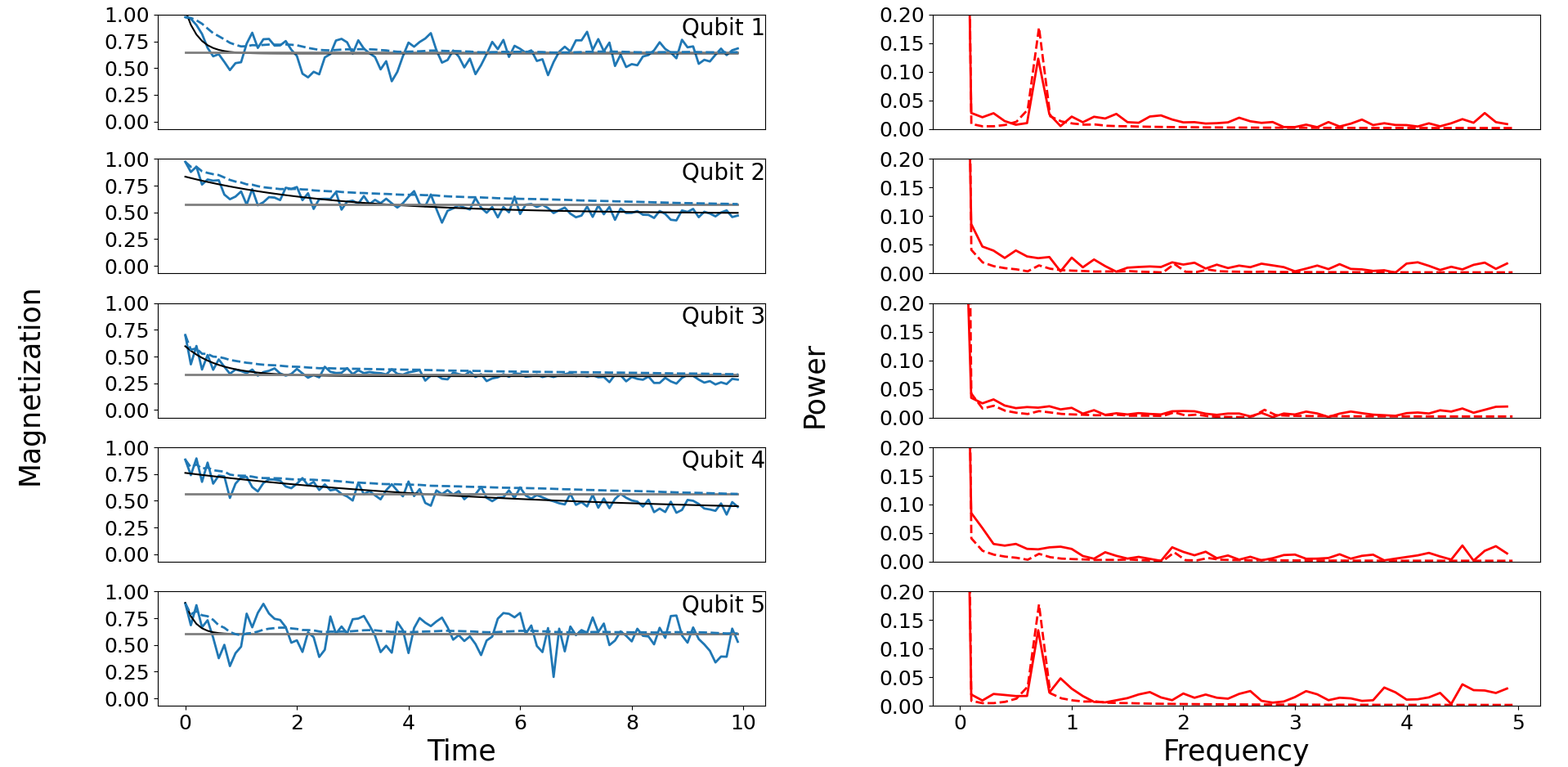}
    \caption{Local dynamics obtained from quantum hardware of the $5$-qubit topological mirror model (Fig. 1(b) of the main text). $J=2$, $J'=4$, $h_x=1$. Left: time-dependent real-space magnetization in the $z$-direction for each qubit plotted with the solid blue line, alongside its local moving average magnetization (dotted blue line) local total average magnetization (gray line) and transient fit (black line). Time is reported in units of the inverse magnetic field $1/h_{x}$. Right: Fourier transformed power spectrum in the frequency space for each qubit. Frequency is reported in units of the magnetic field $h_{x}$. Results are overlaid with its closed system power spectrum in dotted red lines (see next section Sec. \ref{sec:closed_dynamics}.}
    \label{fig:running_2442}
\end{figure}

We comment on the shot noise of our experiments. A total of 8192 shots are used in all quantum hardware results as mentioned above. Our corresponding shot noise is very small compared to the macroscopic features produced by the data; this is confirmed by back-checking with the closed system calculation where its power spectrum is plotted alongside the main result above, with small differences which can be attributed to the noise intrinsic to the quantum hardware. The dotted blue line highlights the overall trend of temporal decay among the bulk modes and consistent oscillations of the surface modes, which can be subtracted from the microscopic oscillations due to the simulation noise.

We remark that, as of the current date, the default amount of shots used in typical IBM quantum experiments is 4000 shots (see IBM  Quantum Administration FAQ), and barring the newly tested experimental processors made available within the last couple of months, the maximum number of allowed shots on almost all devices was previously 8192 (see previous Ref. \cite{ravi2022quantum}).

\section{Closed System Dynamics}\label{sec:closed_dynamics}
In this section we report results obtained by exact diagonalization, which pertain to closed systems that represent our models in an ideal setting. These results serve as benchmarks for the dynamics obtained on quantum hardware, which embody open quantum systems. 
Additionally, exact diagonalization sheds insight into finite-size scaling effects of the systems. We argue that, while constant-depth circuit simulations are constrained by the number of qubits available on the quantum hardware, classical exact diagonalization is constrained by the size of the Hilbert space of the simulated system and available classical computational resources.

\begin{figure}[hbt!]
\centering
    \includegraphics[scale = .4]{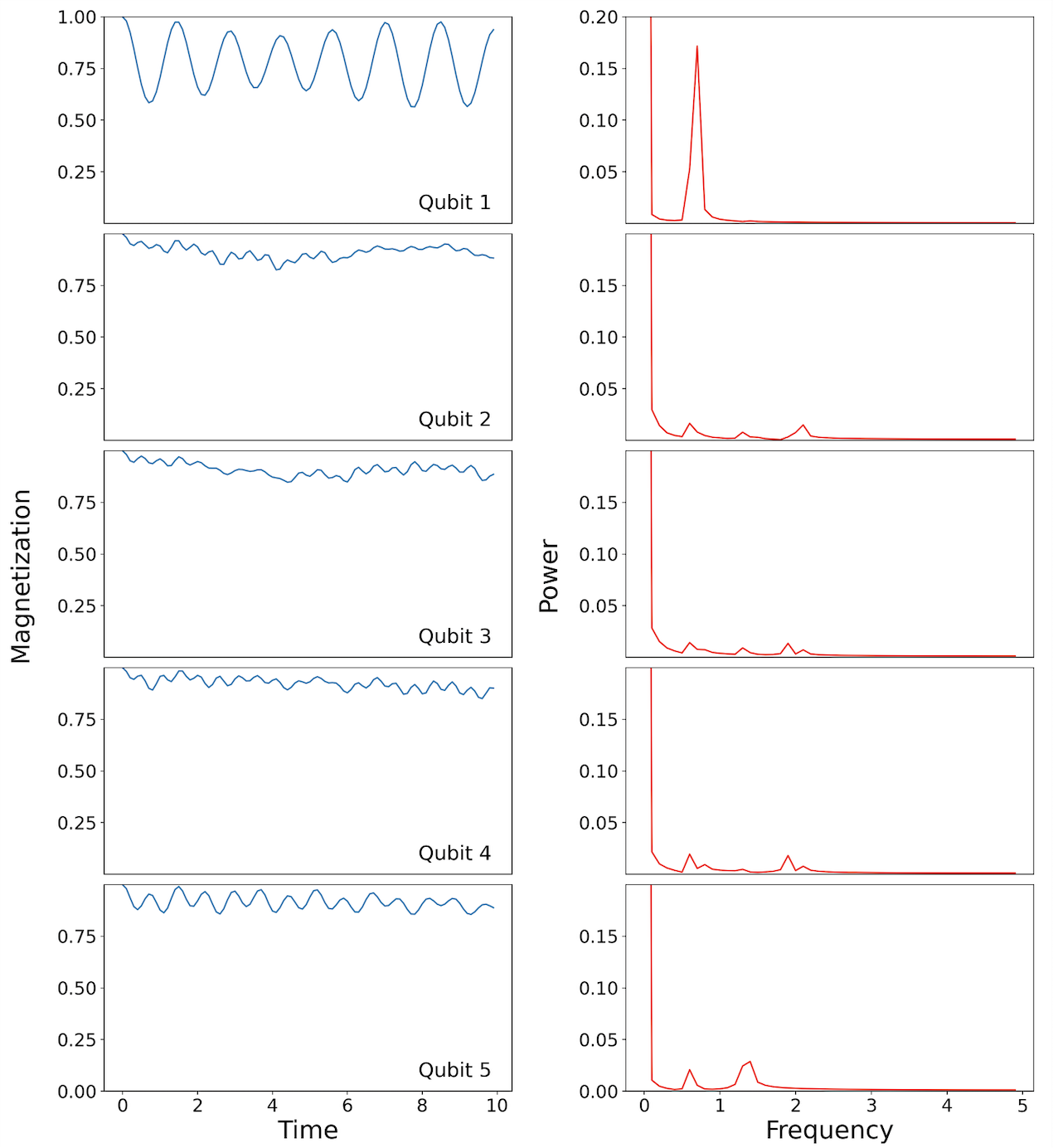}
    \caption{Closed system dynamics of the $5$-qubit staggered coupling model (Fig. 1(a) of the main text). We replicate the same initial conditions as the simulation depicted in Fig. \ref{fig:(4242)_cd}. $J=2$, $J'=4$, $h_x=1$. Left: time-dependent real-space magnetization in the $z$-direction for each qubit. Time is reported in units of the inverse magnetic field $1/h^{x}$. Right: Fourier transformed power spectrum in the frequency space for each qubit. Frequency is reported in units of the magnetic field $h^{x}$.}
    \label{fig:closed_2424}
\end{figure}

\begin{figure}[hbt!]
\centering
    \includegraphics[scale = .4]{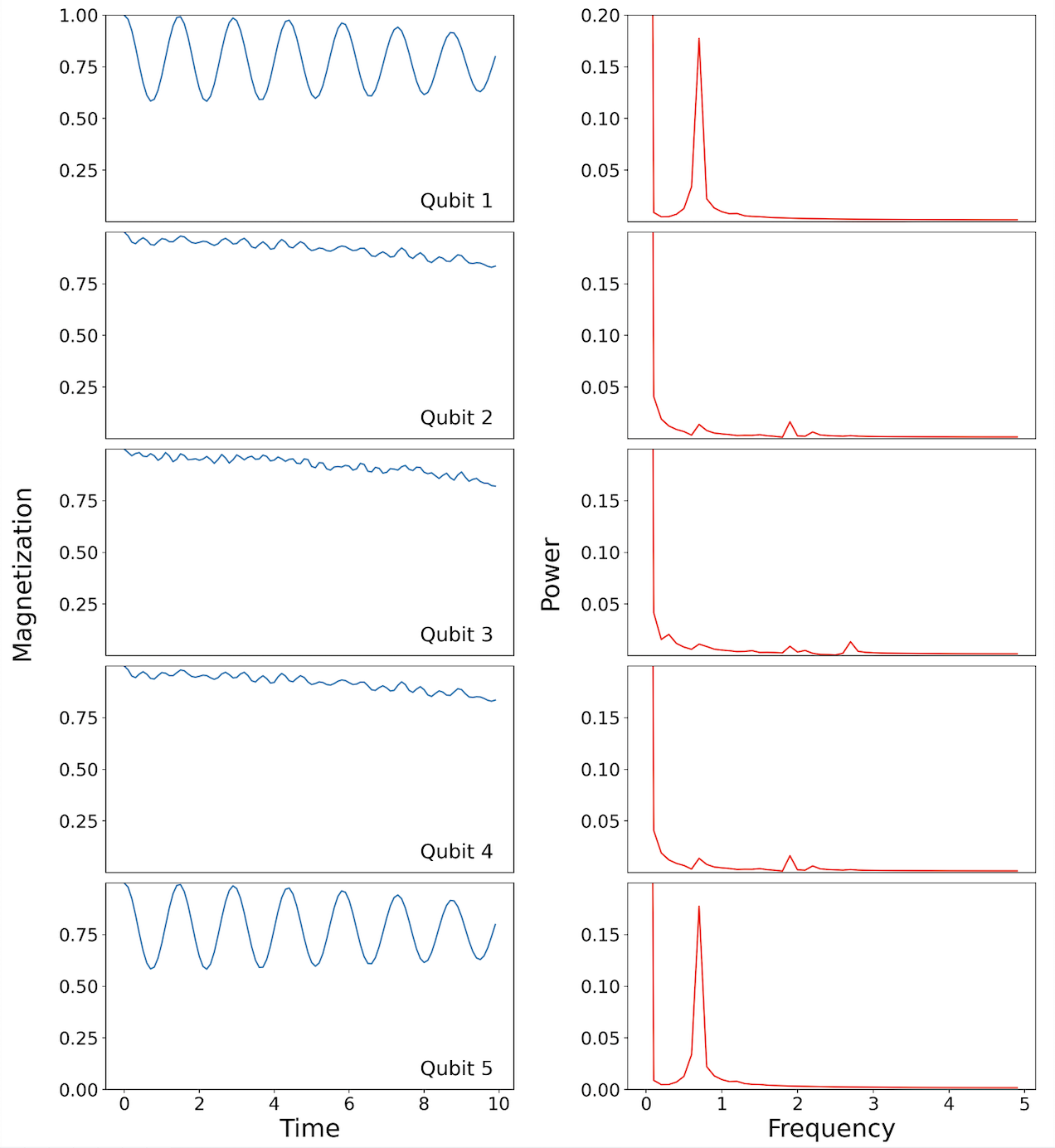}
    \caption{Closed system dynamics of the $5$-qubit topological mirror model (Fig. 1(b) of the main text). We replicate the same initial conditions as the simulation depicted in Fig. 3 of the main text. $J=2$, $J'=4$, $h_x=1$. Left: time-dependent real-space magnetization in the $z$-direction for each qubit. Time is reported in units of the inverse magnetic field $1/h^{x}$. Right: Fourier transformed power spectrum in the frequency space for each qubit. Frequency is reported in units of the magnetic field $h^{x}$.}
    \label{fig:closed_2442}
\end{figure}

\begin{figure}[hbt!]
\centering
    \includegraphics[scale = .4]{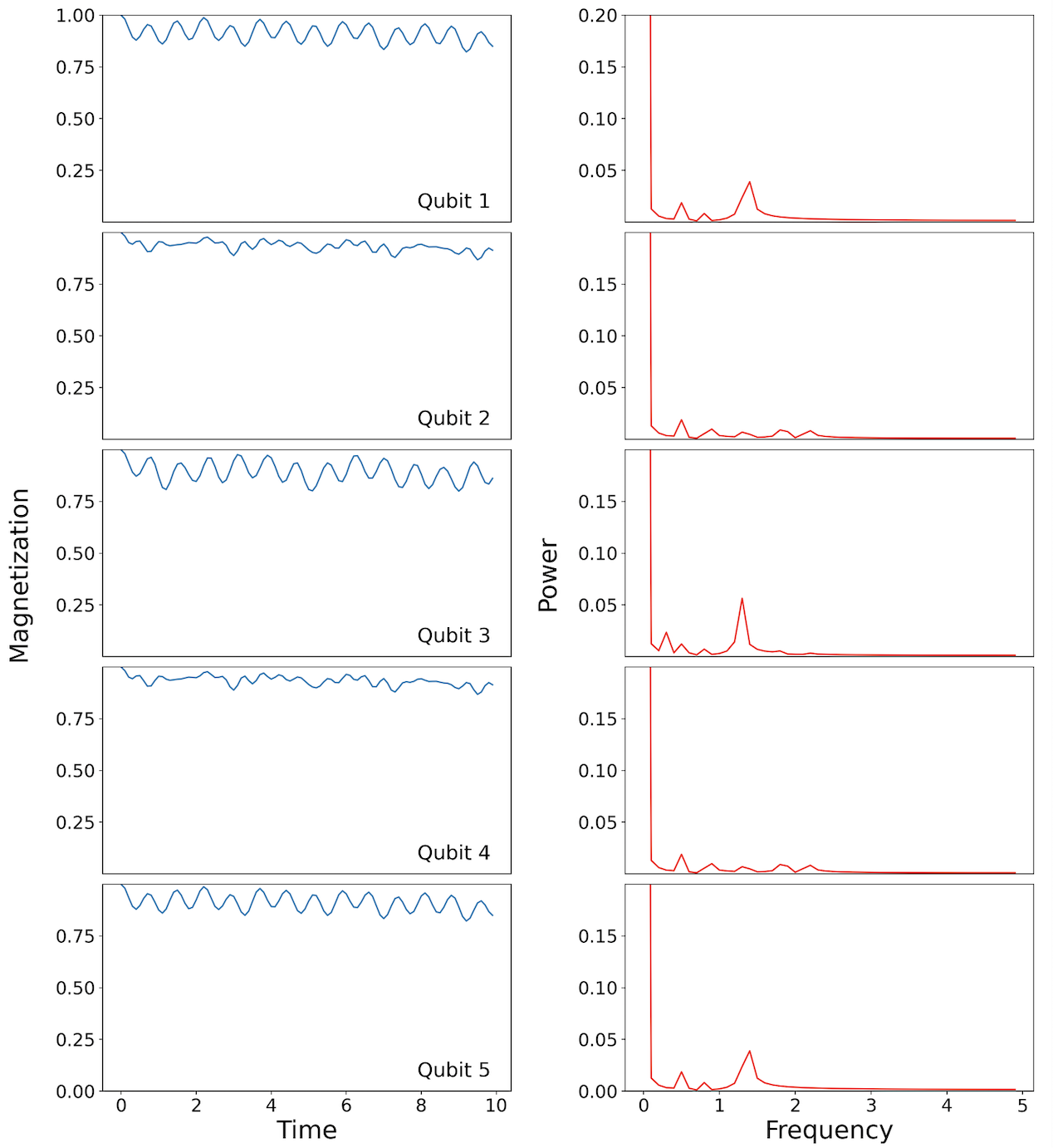}
    \caption{Closed system dynamics of the $5$-qubit topological defect model (Fig. 1(c) of the main text). We replicate the same initial conditions as the simulation depicted in Fig. \ref{fig:(4224)_cd}. $J=2$, $J'=4$, $h_x=1$. Left: time-dependent real-space magnetization in the $z$-direction for each qubit. Time is reported in units of the inverse magnetic field $1/h^{x}$. Right: Fourier transformed power spectrum in the frequency space for each qubit. Frequency is reported in units of the magnetic field $h^{x}$.} 
    \label{fig:closed_4224}
\end{figure}

\begin{figure}[hbt!]
\centering
    \includegraphics[scale = .4]{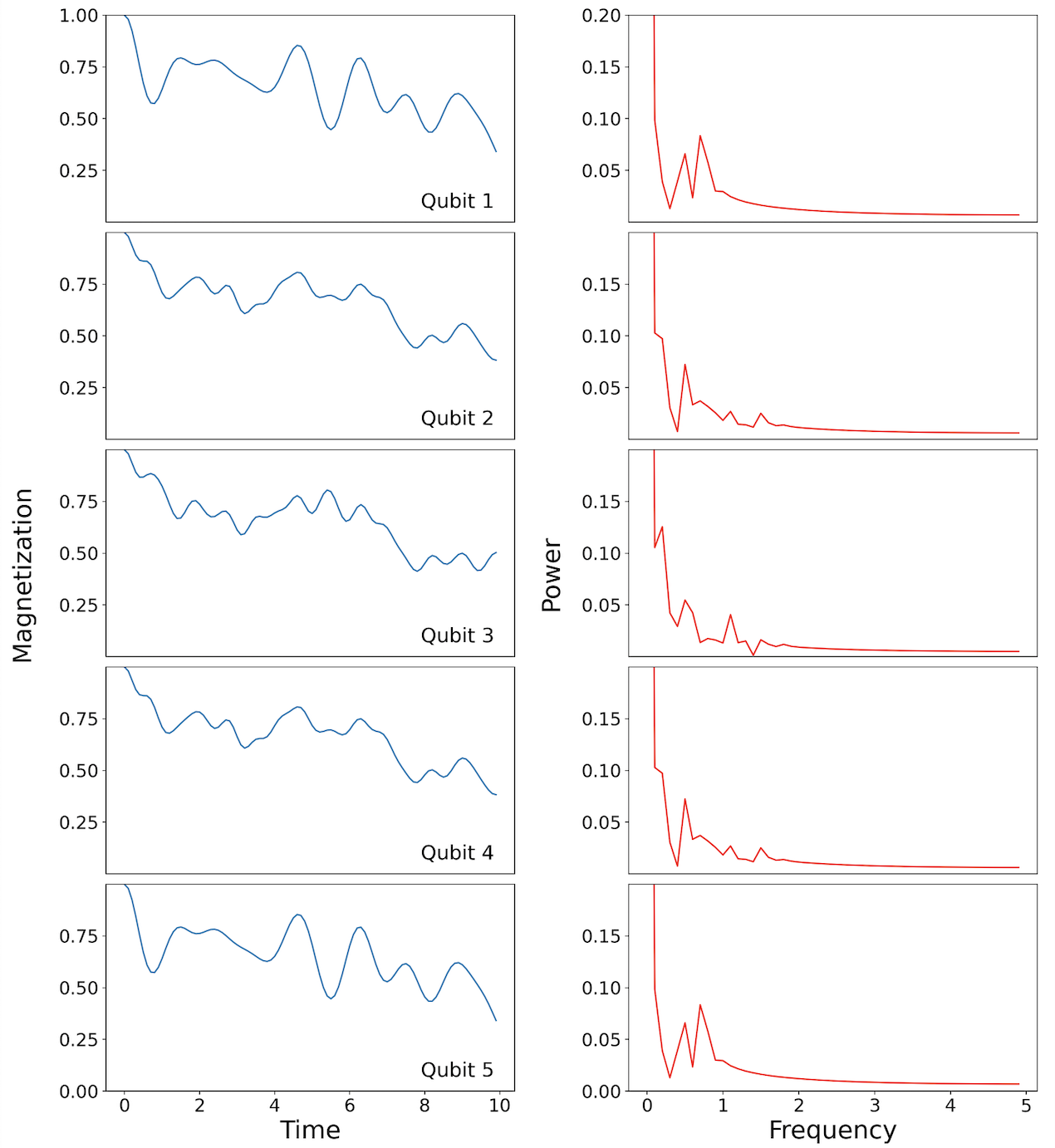}
    \caption{Closed system dynamics of the $5$-qubit uniform coupling model, $J=J'=2$. Time is reported in units of the inverse magnetic field $1/h^{x}$, frequency is reported in units of the magnetic field $h^{x}$. This should be compared to the data in Fig. \ref{fig:(2222)_cd}.}
    \label{fig:closed_2222}
\end{figure}

Figs. \ref{fig:closed_2424}, \ref{fig:closed_2442}, \ref{fig:(4224)_cd} display the closed system dynamics of the models showcased in Fig. 1(a), Fig. 1(b), and Fig. 1(c) of the main text, respectively. Fig. \ref{fig:closed_2222} displays the closed system dynamics of the uniform coupling model.

The closed system analogues of each model exhibit similar behavior and supplement the open system result. Interestingly, we observe noticeably different dynamics in the non-topological $5$-qubit uniform coupling model, where the bulk and boundary are indistinguishable, and the Fourier transformed power spectrum exhibits more pronounced and varied peaks before reaching frequency $= 2$ in units of the applied magnetic field, when compared to the topological models, which notably exhibit power spectrums with patterned features.

\begin{figure} [hbt!]
\centering
    \includegraphics[scale = .4]{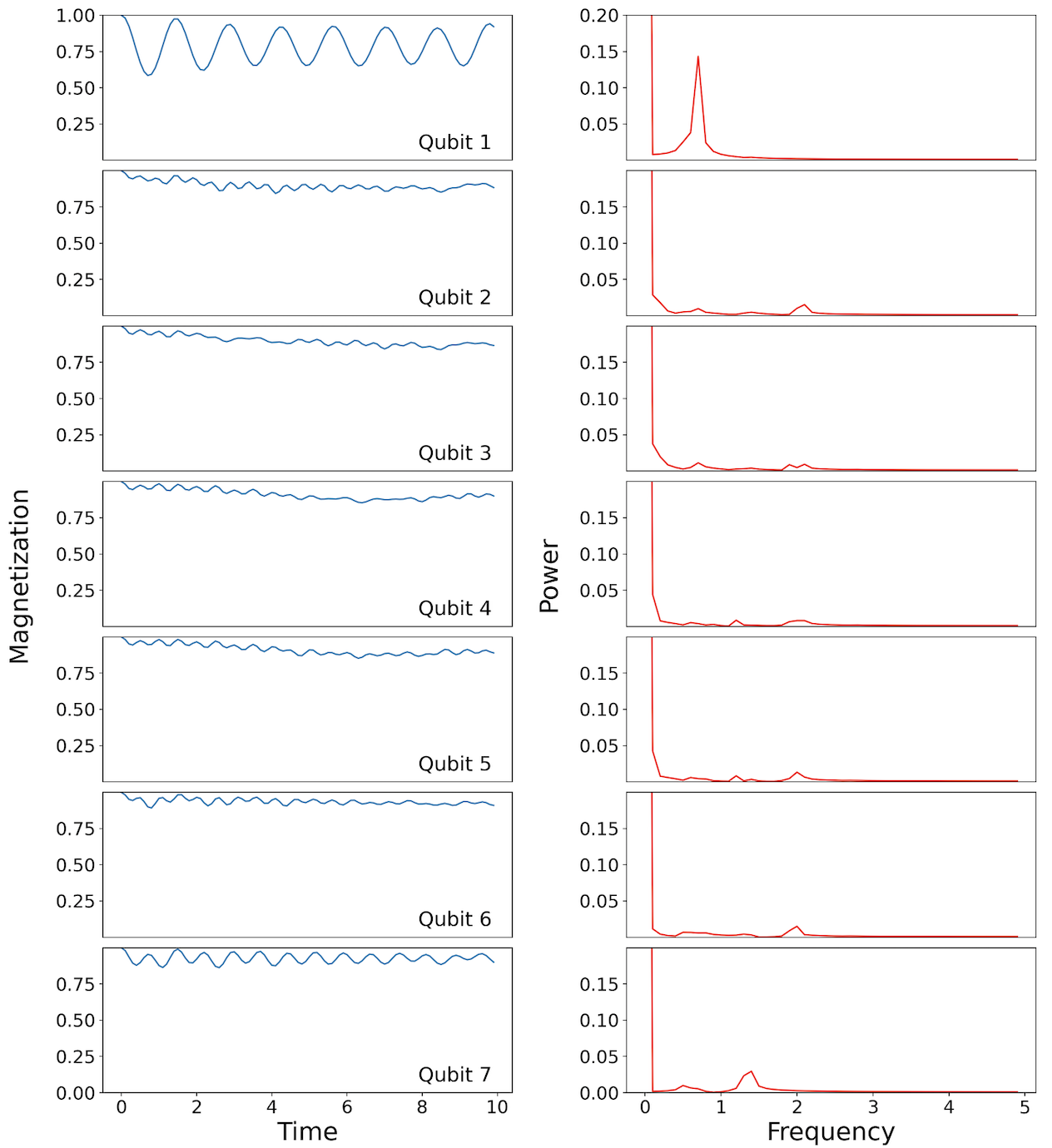}
    \caption{Closed system dynamics of a $7$-qubit staggered coupling chain with alternating couplings $J$ and $J'$. 
    $J=2$, $J'=4$, $h_x=1$. Time is reported in units of the inverse magnetic field $1/h^{x}$, frequency is reported in units of the magnetic field $h^{x}$. Since the system has an odd number of sites, one topological mode is visible on the weakly-coupled edge qubit 1.}\label{fig:closed_242424}
\end{figure}

\begin{figure} [hbt!]
\centering
    \includegraphics[scale = .4]{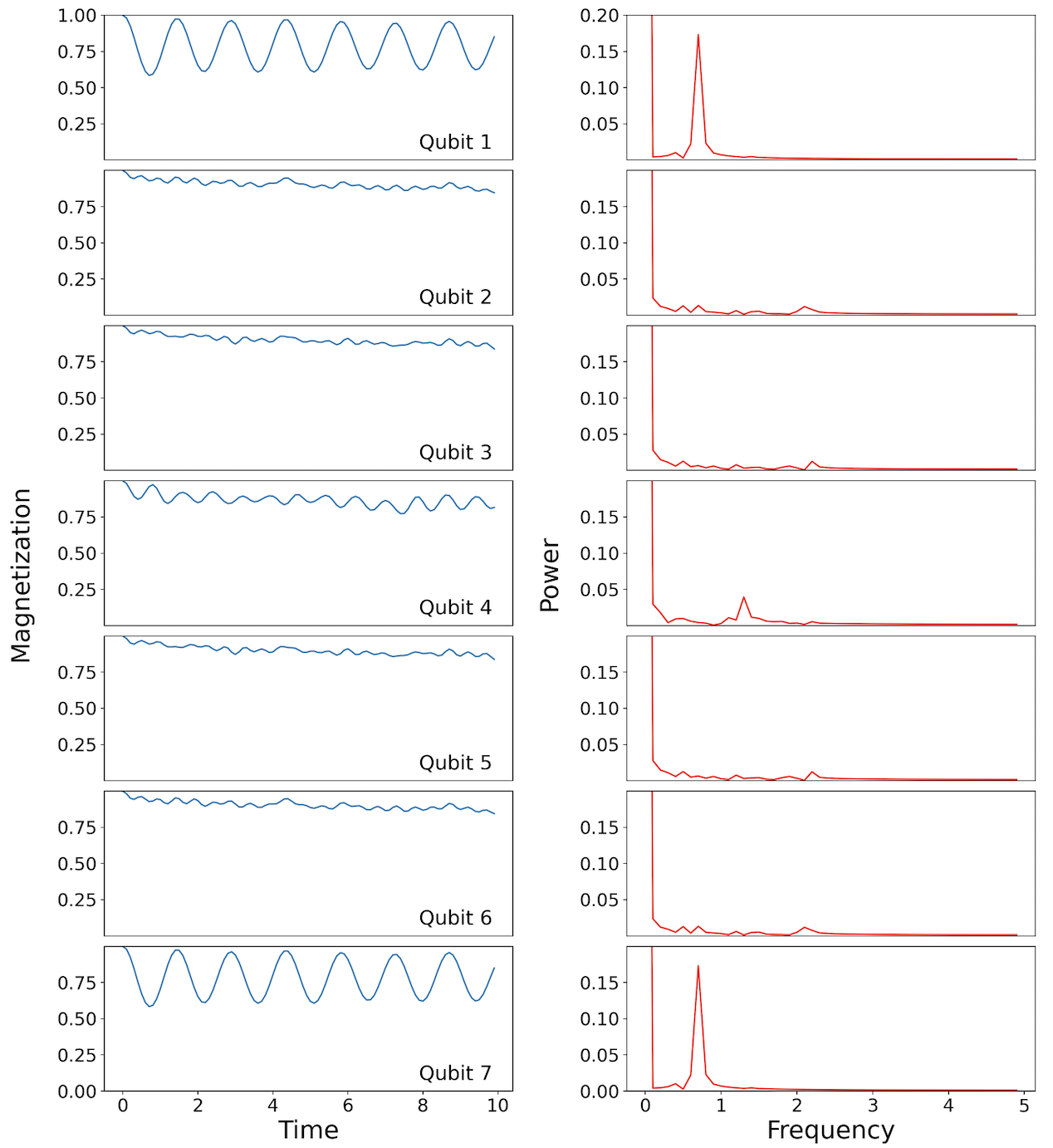}
    \caption{Closed system dynamics of a $7$-qubit topological defect system with weak coupling in the center, specified by $J_{1} = 2$, $J_{2} = 4$, $J_{3} = 2$, $J_{4} = 2$, $J_{5} = 4$, $J_{6} = 2$, $h_x=1$. Time is reported in units of the inverse magnetic field $1/h^{x}$, frequency is reported in units of the magnetic field $h^{x}$.}\label{fig:closed_242242}
\end{figure}

\begin{figure} [hbt!]
\centering
    \includegraphics[scale = .4]{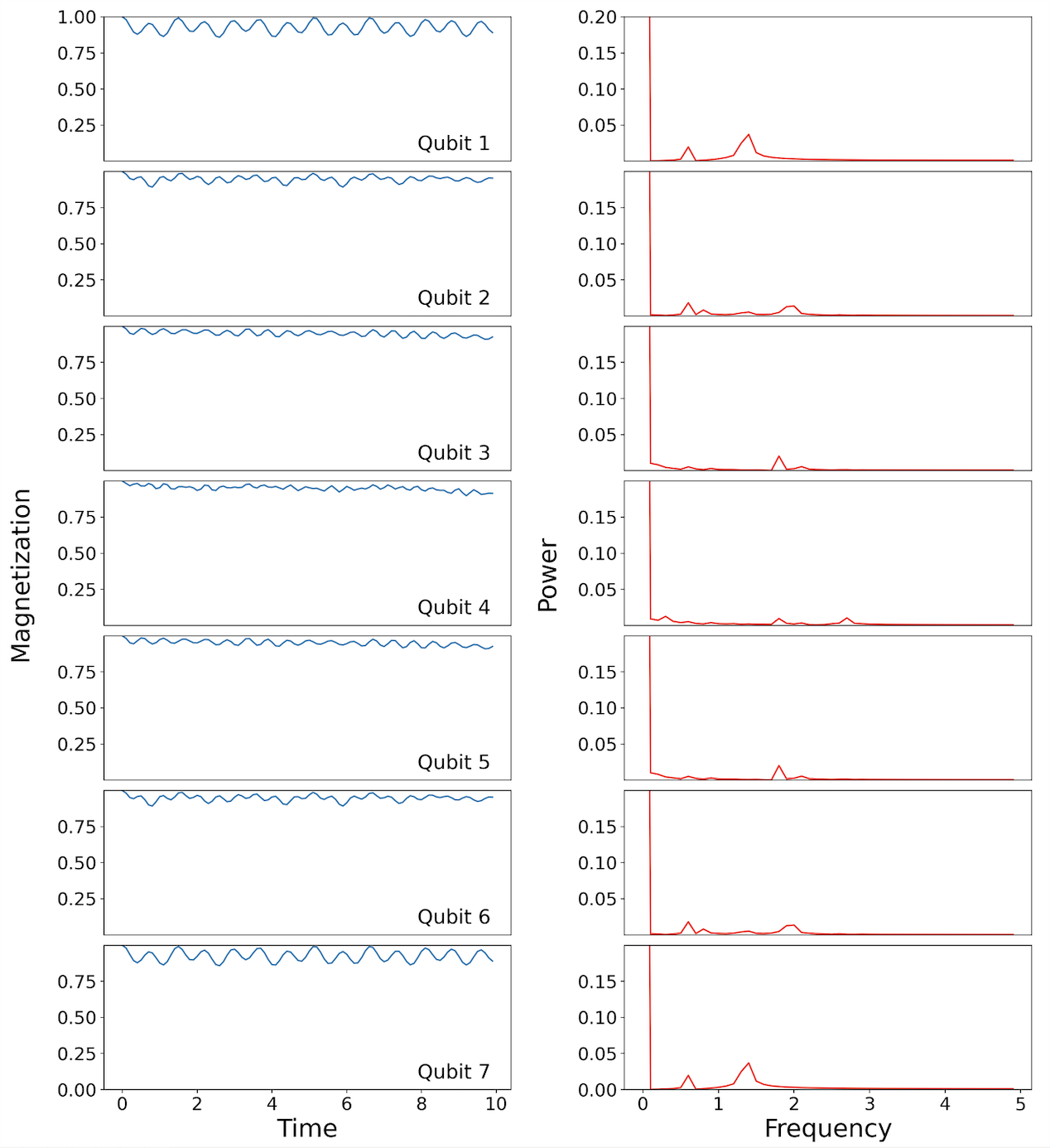}
    \caption{Closed system dynamics of a $7$-qubit topological defect system with strong coupling in the center, specified by $J_{1} = 4$, $J_{2} = 2$, $J_{3} = 4$, $J_{4} = 4$, $J_{5} = 2$, $J_{6} = 4$, $h=1$. Time is reported in units of the inverse magnetic field $1/h^{x}$, frequency is reported in units of the magnetic field $h^{x}$.}\label{fig:closed_424424}
\end{figure}

Additionally, we provide the closed system dynamics for three topological $7$-qubit systems and a uniform $7$-qubit chain. We accomplish this by employing the same Hamiltonian, $\mathcal{H} = - \sum_{i=1}^{N-1} J_{z, i} Z_{i} Z_{i+1} -  h_{x}\sum_{i=1}^{N} X_{i}$, with $N$ = 7 spins. 
Fig. \ref{fig:closed_242424} pertains to the staggered coupling model extended to 7 sites, inspired by the SSH model \cite{PhysRevLett.42.1698}. 
Next, we present two systems which analogue the topological defect model, created by imprinting a topological defect in the center (qubit 4) between two topological configurations. In Fig. \ref{fig:closed_242242}, we weakly link two spin chains consisting of weakly coupled outer qubits and strongly coupled inner qubits. In Fig. \ref{fig:closed_424424}, we implement the opposite by strongly linking two spin chains consisting of strongly coupled outer qubits and weakly coupled inner qubits.

\begin{figure} [hbt!]
\centering
    \includegraphics[scale = .4]{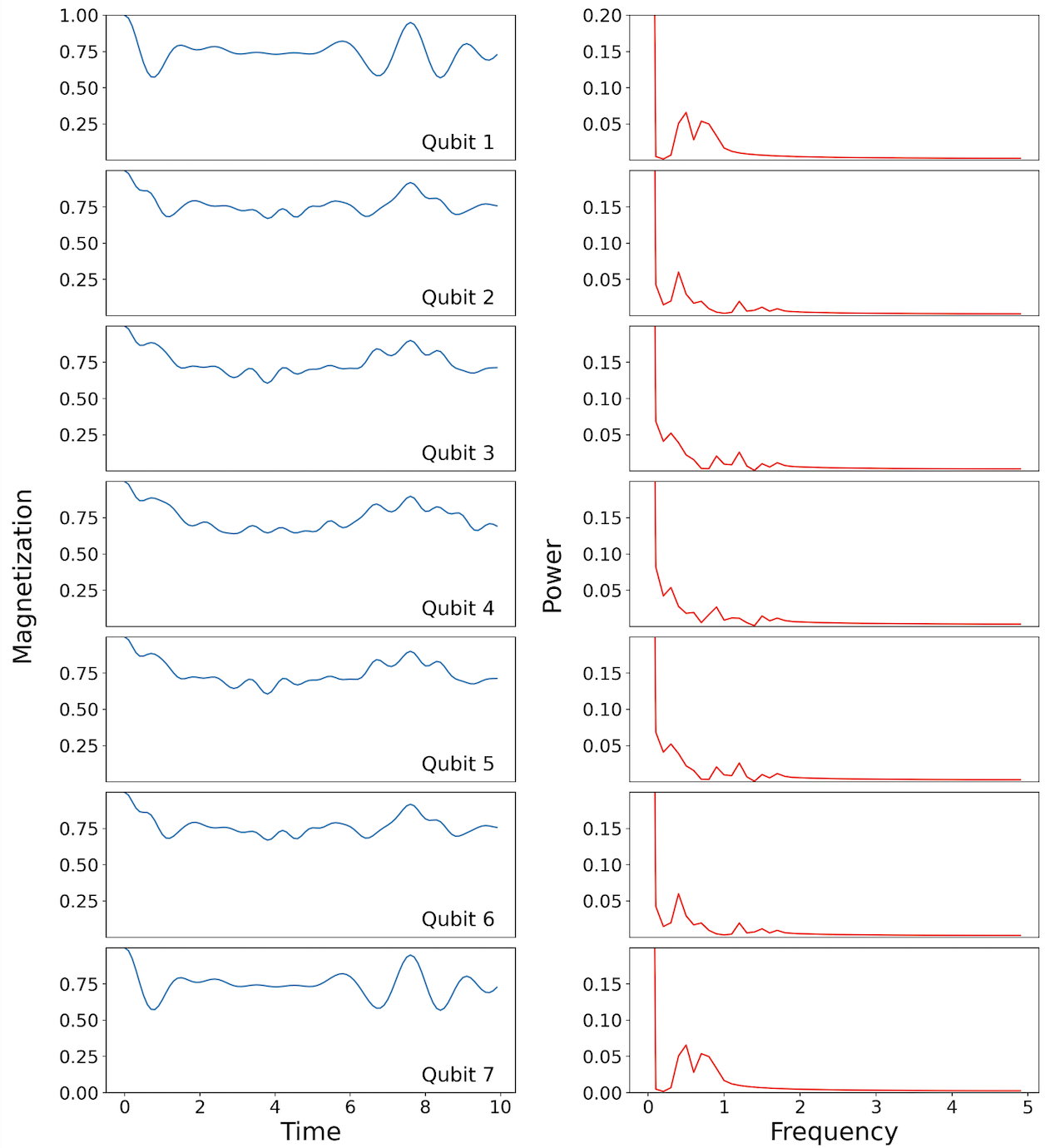}
    \caption{Closed system dynamics of a $7$-qubit uniform chain, $J=J'=2$, $h=1$. Time is reported in units of the inverse magnetic field $1/h^{x}$, frequency is reported in units of the magnetic field $h^{x}$.}\label{fig:closed_222222}
\end{figure}

Bulk-boundary separation affects the obtained results, as discussed in Sec. \ref{sec:additional}. The weakly linked 7-qubit topological defect model of Fig. \ref{fig:closed_242242} analogues the 5-qubit 'trimer' defect model discussed in Fig. \ref{fig:(4224)_cd}. For the 7-qubit chain, larger separation between the system bulk and boundary creates dimer pairs with higher localization at the edges of the lattice.


\clearpage
\bibliography{supp}